\documentclass[reprint,twocolumn,showpacs,preprintnumbers,amsmath,amssymb,amsfonts,final, a4paper,aps, 
citeautoscript,footinbib,prb,superscriptaddress]{revtex4-1}
\usepackage[pdftex]{graphicx,color}
\usepackage{epstopdf}
\usepackage[latin1]{inputenc}
\usepackage{mathrsfs}
\usepackage{times}
\usepackage{float}
\usepackage[vcentermath]{youngtab}
\usepackage[sort&compress]{natbib}
\usepackage[colorlinks,bookmarks=true,citecolor=blue,linkcolor=blue,urlcolor=blue, breaklinks=true]{hyperref}
\graphicspath{{./figures/}}
\newcommand{\be}{\mathrm{e}}
\begin{document}
%%%%%%%%%%%%%%%%%%%%%%%%%%%%%%%%%%%%%%%%%%%%%%%%%%%%%%%
\title{Symmetry-protected topological order in SU({\em N}) Heisenberg magnets \\
-- quantum entanglement and non-local order parameters}
%%%%%%%%%%%%%%%%%%%%%%%%%%%%%%%%%%%%%%%%%%%%%%%%%%%%%%%
\author{K. Tanimoto}
\affiliation{Division of Physics and Astronomy, Graduate School of Science, Kyoto University, 
Kitashirakawa Oiwake-Cho, Kyoto 606-8502, Japan.}
\affiliation{Yukawa Institute for Theoretical Physics, 
Kyoto University, Kitashirakawa Oiwake-Cho, Kyoto 606-8502, Japan.}
\author{K. Totsuka}
\affiliation{Yukawa Institute for Theoretical Physics, 
Kyoto University, Kitashirakawa Oiwake-Cho, Kyoto 606-8502, Japan.}
%%%%%%%%%%%%%%%%%%%%%%%%%%%%%%%%%%%%%%%%%%%%%%%%%%%%%%%
\date{\today}
\pacs{75.10.Pq, 71.10.Pm}

% PACS used:
% 71.10.Pm Fermions in reduced dimensions 
% 75.10.Pq Spin chain models
%%%%%%%%%%%%%%%%%%%%%%%%%%%%%%%%%%%%%%%%%%%%%%%%%%%%%%%
\begin{abstract}
In this paper, we investigate topological properties of the ground state of the SU($N$) Heisenberg chain,    
which is argued to be relevant to the Mott-insulating phase of alkaline-earth cold fermions in a one-dimensional 
optical lattice.   By calculating the entanglement spectrum, we show that the ground state is in one of 
the topological phases protected by SU($N$) symmetry.  We then discuss an alternative characterization of it 
with non-local string order parameters.  We also consider how the reduction of the protecting symmetry affects 
the topological phase paying particular attention to the entanglement spectrum.  
\end{abstract}
%%%%%%%%%%%%%%%%%%%%%%%%%%%%%%%%%%%%%%%%%%%%%%%%%%%%%%%
\maketitle
%%%%%%%%%%%%%%%%%%%%%%%%%%%%%%%%%%%%%%%%%%%%%%%%%%%%%%%
\section{Introduction}
%%%%%%%%%%%%%%%%%%%%%%%%%%%%%%%%%%%%%%%%%%%%%%%%%%%%%%%
Symmety in physics not only is the key to understanding phases of matter 
but also play a vital role in unifying seemingly different things and uncovering 
fundamental principles underlying them.   
In particular, unitary groups have been playing very important roles in quantum mechanics 
as the orthogonal groups in classical mechanics. 
For instance, SU(3) is the fundamental symmetry underlying the quantum chromodynamics (QCD) 
of strong interactions.  
In traditional condensed-matter physics, however, high symmetry like SU($N$) is usually realized,  
aside from few exceptions, only in rather idealized situations and has been mainly used 
as mathematical convenience that makes problems tractable.  
For instance, in the large-$N$ approximations, we replace the physical symmetry SU(2) with 
SU($N$) and use $1/N$ as the (small) control parameter of the approximation hoping that there is 
a smooth crossover down to $N=2$.   
 
Recent suggestions\cite{Cazalilla-H-U-09,Gorshkov-et-al-10} that SU($N$)-symmetric fermion systems could be 
simulated using the alkaline-earth atoms and their cousins (${}^{171}$Yb, ${}^{173}$Yb, ${}^{87}$Sr, etc.) 
loaded in optical lattices opened a new era of SU($N$) physics 
\cite{Kitagawa-et-al-PRA-08,DeSalvo-Y-M-M-K-10} 
(see, e.g., Refs.~\onlinecite{Sugawa-T-E-T-Yb-review-13,Cazalilla-R-14} for recent reviews).  
For instance, the SU($N$) generalization of quantum magnetism is of direct relevance to the Mott-insulating 
regime of these systems.  
The SU($N$) ``spin'' models provide us with examples of underconstrained 
systems that yield, on top of usual ``magnetically ordered'' states, various unconventional states, e.g., 
deconfined criticalities\cite{Kaul-S-12,Harada-S-O-M-L-W-T-K-13}, 
an algebraic spin liquid\cite{Corboz-L-L-P-M-12} and a chiral spin liquid\cite{Hermele-G-11}. 

On the other hand, topological states of matter\cite{Wen-book-04} have been subjects of extensive research 
for the past decade.  
Since the advent of topological insulators and superconductors\cite{Qi-Z-RMP-11}, 
it has been widely realized that 
there exists a special class of ``topological'' phases that is stable {\em only} in the presence of 
certain symmetries\cite{Gu-W-09,Pollmann-T-B-O-10,Chen-G-W-11,Chen-G-L-W-12,Vishwanath-S-13}.  
This class of topological phases is called ``symmetry-protected topological (SPT)''\cite{Gu-W-09} 
as it is topologically protected only when we impose symmetries on the system in question, 
and otherwise they reduce to trivial ones.  
The catalogue of possible topological phases depends crucially on the symmetry 
we impose and different lists of possible phases may be obtained for different protecting symmetries 
(see, e.g., Ref.~\onlinecite{Chen-G-L-W-13} for a catalogue of SPT phases).  
One defining property of SPT phases is the existence of gapless boundary excitations 
(edge states) that are intrinsically different from those in the gapped bulk.  
A modern mathematical way of observing the edge states would be to use 
the entanglement spectrum\cite{Li-H-08} that is obtained solely from the ground-state wave function.  
In the following, we heavily use the entanglement spectrum in characterizing topological phases.  

Despite the recent effort\cite{Chen-G-W-11} in systematically enumerating possible SPT phases in one dimension, 
not much is known, except for a few examples, about how to observe those phases in realistic settings. 
Recently, it has been suggested\cite{Nonne-M-C-L-T-13,Bois-C-L-M-T-15} that 
a class of SPT phases is realized in the Mott-insulating region of the alkaline-earth cold fermions,  
and this is one of the motivations of our study here.  
Specifically, deep inside the Mott phase at half-filling, the low-energy physics of a system 
of alkaline-earth fermions is described by an SU($N$) ``spin'' model 
(see Secs.~\ref{sec:relation-to-AE} and \ref{sec:strong-coupling}) whose ground state is expected 
to be in one of the topological phases predicted in Ref.~\onlinecite{Duivenvoorden-Q-13}.   
Therefore the alkaline-earth fermions provide us with a unique arena for the realization of new SPT phases 
in a very controlled manner. 
Our goal is to clarify the nature of the ground state of the above SU($N$) spin Hamiltonian in several complementary 
ways and demonstrate the use of non-local string order parameters to detect the phase.  

The outline of this paper is as follows.  
In Sec.~\ref{sec:model}, we introduce the SU($N$) Heisenberg model and sketch how it is derived 
as the effective Hamiltonian for the Mott-insulating phase of the alkaline-earth cold fermions on 
a one-dimensional optical lattice.  A variant of the Heisenberg model that gives useful insights about 
the topological properties of the original model is introduced as well.  
After briefly summarizing the minimal background of SPT phases expected for our SU($N$) spin systems, 
we try, in Sec.~\ref{sec:SPT}, to characterize the topological properties of the ground state 
of the SU($N$) Heisenberg model using its entanglement spectrum. 
By carefully investigating the structure of the spectrum obtained for $N=4$, 
we present a strong evidence that the ground state of the SU(4) Heisenberg model is in one of the SU(4) 
topological phases.  In Sec.~\ref{sec:string-orer-parameter}, we present an alternative way of characterizing the SU($N$) 
SPT phases using {\em non-local} string order parameters.  

Although the alkaline-earth fermions, that motivated our study, possess very precise SU($N$) symmetry, 
it would be interesting theoretically to consider the situations where the original SU($N$) symmetry 
gets lowered.  We investigate this problem in Sec.~\ref{sec:symmetry-reduction} to find that,  
depending on $N$, the system remains topological even after the SU($N$) symmetry is relaxed.   
Summary of the main results is given in Sec.~\ref{sec:summary}.   
%%%%%%%%%%%%%%%%%%%%%%%%%%%%%%%%%%%%%%%%%%%%%%%%%%%%%%%
\section{Model} 
\label{sec:model}
%%%%%%%%%%%%%%%%%%%%%%%%%%%%%%%%%%%%%%%%%%%%%%%%%%%%%%%
In this paper, we consider the ground-state properties of the following Hamiltonian
\begin{equation}
\mathcal{H}_{\text{Heis}} = \mathcal{J} \sum_{A=1}^{N^{2}-1} 
\mathcal{S}_{i}^{A} \mathcal{S}_{i+1}^{A} 
\label{eqn:SUN-Heisenberg}
\end{equation}
where $\mathcal{S}_{i}^{A}$ ($A=1,\ldots, N^{2}-1$) denote the SU($N$) generators.  
In SU($N$), instead of fixing spin $S$, one has to specify the irreducible representation(s) to which 
the generators $\mathcal{S}_{i}^{A}$ belong. 
In the following,  $\mathcal{S}_{i}^{A}$ ($A=1,\ldots, N^{2}-1$) denote, unless otherwise stated, 
the SU($N$) generators in the irreducible representation characterized by the following Young diagram 
with $N/2$ rows and two columns: 
\begin{equation}
\text{\scriptsize $N/2$} \left\{ 
\yng(2,2,2)
 \right.  \quad (N=\text{even}) \; .
\label{eqn:Young-diagram-GS}
\end{equation}

It is well-known that the low-energy physics of the SU($N$) Heisenberg model depends 
crucially on the representation(s) we put on the individual lattice sites.  
For the fully-symmetrized representation ${\tiny \yng(2)\cdots \yng(1)}$ 
($n_{\text{c}}$ boxes),  
the exact Bethe-ansatz solutions are available\cite{Sutherland-75,Andrei-J-84,Johannesson-86}; 
the ground state is known to be gapless and described by the level-$n_{\text{c}}$  
SU($N$) Wess-Zumino-Witten conformal field theory with the central charge 
$c=n_{\text{c}}(N^{2}-1)/(N+n_{\text{c}})$\cite{Alcaraz-M-SUN-89}.   
For {\em any} translationally invariant choice of representations (i.e., the same representation is 
assigned on every site), we can show that the SU($N$) chain, which has a unique (finite-size) ground 
state\footnote{%
The proof of the existence of low-lying states works regardless of whether the ground state 
is unique or not.  However, unless the (finite-size) ground state is unique, the proof does not 
imply anything about {\em excited} states.}, 
is either gapless or has degenerate ground states (with broken symmetries) provided that 
the number of boxes $n_{\text{Y}}$ in the Young diagram is {\em not} divisible by $N$\cite{Affleck-L-86}.   
In other words, except for the cases of $n_{\text{Y}}=0$ (mod $N$) 
[including the one shown in Eq.~\eqref{eqn:Young-diagram-GS} 
which is relevant to our spin chain], this statement excludes the possibility of gapped topological 
ground states.  
Remarkably, this is perfectly consistent with the recent group-cohomology classification of 
the gapped SPT phases\cite{Duivenvoorden-Q-13} (see Sec.~\ref{sec:SUN-SPT} for the detail).  
There is also an attempt\cite{Greiter-R-07} 
at summarizing these observations into a ``generalized'' Haldane conjecture.   

Some insights about the  nature of the ground state of \eqref{eqn:SUN-Heisenberg} are gained 
from the large-$N$ analysis\cite{Marston-A-89,Read-S-NP-89,Read-S-90} as well.  
For $N/2$ rows but with a single column, 
the ground state is expected dimerized\cite{Marston-A-89}, while, for two columns, we may have 
a gapped translationally invariant ground state\cite{Read-S-NP-89,Read-S-90}, which we will argue 
to be topological.   

%%%%%%%%%%%%%%%%%%%%%%%%%%%%%%%%%%%%%%%%%%%%%%%%%%%%%%%
\subsection{Relation to cold fermion systems}
\label{sec:relation-to-AE}
%%%%%%%%%%%%%%%%%%%%%%%%%%%%%%%%%%%%%%%%%%%%%%%%%%%%%%%
It has been argued in Refs.~\onlinecite{Nonne-M-C-L-T-13,Bois-C-L-M-T-15} that 
the Hamiltonian \eqref{eqn:SUN-Heisenberg} emerges as the effective Hamiltonian 
in the Mott-insulating region of the alkaline-earth cold fermions loaded in a one-dimensional 
optical lattice at half-filling.   
To emphasize the relevance of our results to experimentally realizable systems, 
we sketch how the model $\mathcal{H}_{\text{Heis}}$ is derived from the cold-fermion systems 
in the Mott region. 

It is known that the decoupling between the nuclear spin ($I$) and the total electron angular momentum 
makes it possible to organize the $(2I+1)$ nuclear-spin states of each atom into a multiplet of larger 
SU($2I+1$)-symmetry.  
Specifically, the interaction between two like alkaline-earth atoms does {\em not} depend on 
the nuclear-spin states of each and hence is SU($2I+1$)-symmetric\cite{Cazalilla-H-U-09,Gorshkov-et-al-10}.  
Moreover, one can add one more degree of freedom ({\em orbital}) by taking into account 
the first meta-stable excited states (in $^{3}P_{0}$; denoted as ``$e$'') 
as well as the atomic ground state in $^{1}S_{0}$ (``$g$'').\footnote{%
A remark is in order here about the use of the terminology `orbital' here. 
In the case of electrons in crystals, orbital is closely tied to the spatial structure of the wave function 
and often allows pair-hopping processes that break continuous orbital symmetry down to a discrete 
one.  The two orbitals $g$ and $e$, on the other hand, are internal degrees of freedom and, 
in the absence of the internal conversion between $g$ and $e$, 
the system retains at least orbital U(1) symmetry.}  
That this SU($2I+1$)-symmetry holds for both orbitals with very high accuracy has been verified in 
recent scattering-length measurements\cite{Kitagawa-et-al-PRA-08,Zhang-et-al-Sr-14,Scazza-et-al-14}.   

When loaded into a one-dimensional optical lattice, the system of alkaline-earth cold fermions 
is described by the following Hubbard-like Hamiltonian\cite{Gorshkov-et-al-10}
\begin{equation}
\begin{split}
  \mathcal{H}_{\text{G}} =& 
  -  \sum_{i}\sum_{m=g,e} t^{(m)} \sum_{\alpha=1}^{N} 
   \left(c_{m\alpha,\,i}^\dag c_{m\alpha,\,i+1}  + \text{h.c.}\right) \\
&  -\sum_{m=g,e}\mu^{(m)}_{\text{G}} \sum_i n_{m,i}  
 +\sum_{i}\sum_{m=g,e} \frac{U^{(m)}_{\text{G}}}{2} n_{m,\,i}(n_{m,\,i}-1) \\
&  +V_{\text{G}} \sum_i n_{g,\,i} n_{e,\,i} 
  + V_{\text{ex}}^{g\text{-}e} \sum_{i,\alpha \beta} 
  c_{g\alpha,\,i}^\dag c_{e\beta,\,i}^\dag 
  c_{g\beta ,\,i} c_{e\alpha,\,i} ,
  \end{split}
  \label{eqn:Gorshkov-Ham}
\end{equation} 
where $N=2I+1$ denotes the number of nuclear-spin states and 
the operator $c_{m\alpha,\,i}^{\dag}$ creates an atom in the internal state $(\alpha,m)$ 
($\alpha=1,\ldots,N$, $m=g,e$) at the site $i$. 
The number operators are defined as 
$n_{m\alpha,\,i} = c_{m\alpha,\,i}^{\dag}c_{m\alpha,\,i}$ and $n_{m,\,i} = \sum_{\alpha = 1}^N n_{m\alpha,\,i}$. 
As the two orbitals are not symmetry-related, the hopping amplitudes $t^{(m)}$ ($m=g,e$), 
the chemical potential $\mu^{(m)}_{\text{G}}$, and the intra-orbital interaction $U^{(m)}_{\text{G}}$ in general are 
different for the two orbitals.  
The inter-orbital exchange (or, Hund coupling) $V_{\text{ex}}^{g\text{-}e}$ is crucial in determining 
the nature of the Mott-insulating phases\cite{Bois-C-L-M-T-15}.  

Clearly, the Hamiltonian \eqref{eqn:Gorshkov-Ham} is invariant under the SU($N$) transformation 
\begin{equation}
c_{m\alpha,i} \to \sum_{\beta=1}^{N} \mathcal{U}_{\alpha\beta} c_{m\beta,i} 
\quad [\, \mathcal{U} \in \text{SU($N$)} \, ] 
\end{equation}
as well as the multiplication of a global U(1) phase:
\begin{equation}
c_{m\alpha,i} \to \be^{i \theta} c_{m\alpha,i} \; .
\label{eqn:U1-gauge}
\end{equation} 
Borrowing a terminology from the electron systems, we call, in the rest of this paper, 
the degree of freedom associated with \eqref{eqn:U1-gauge} 
``charge'', although the fermions $c_{m\alpha,\,i}$ are charge-neutral in the cold-atom context.  
This and the related systems have been investigated extensively both for SU(2)%
\cite{Nonne-B-C-L-10,Nonne-B-C-L-11,Kobayashi-O-O-Y-M-12,Kobayashi-O-O-Y-M-14} 
and for SU($N$)\cite{Nonne-M-C-L-T-13,Bois-C-L-M-T-15,Szirmai-13}.   
%%%%%%%%%%%%%%%%%%%%%%%%%%%%%%%%%%%%%%%%%%%%%%%%%%%%%%%%
\subsection{Strong-coupling limit}
\label{sec:strong-coupling}
%%%%%%%%%%%%%%%%%%%%%%%%%%%%%%%%%%%%%%%%%%%%%%%%%%%%%%%%
Recently, it has been argued\cite{Nonne-M-C-L-T-13,Bois-C-L-M-T-15} that 
for large positive $U^{(m)}_{\text{G}}$ and $V_{\text{ex}}^{g\text{-}e}$, 
there exists a topological Mott phase protected 
by SU(4)-symmetry.\footnote{To be precise, the protecting symmetry is not SU(4) but 
$\text{PSU(4)}\simeq \text{SU(4)}/\mathbb{Z}_{4}$.}
In order to consider the Mott-insulating phases, it is convenient to start from 
the strong-coupling limit $U_{\text{G}}^{(m)}, V_{\text{G}}, V_{\text{ex}}^{g\text{-}e} \ll t^{(m)}$.  
In this limit, charge fluctuations are strongly suppressed and the SU($N$) ``spin'' and orbital dominate 
the low-energy physics.   
One may introduce the psuedo-spin operator 
$T_i^a = \frac{1}{2}\sum_{\alpha,\beta,m}c_{m\alpha,i}^{\dagger}\sigma_{\alpha\beta}^ac_{m\beta, i}$ 
($a=x,y,z$) for each orbital to rewrite the single-site (i.e., $t^{(m)}=0$) part of the Hamiltonian as
\begin{equation}
\begin{split}
&  \mathcal{H}_{\text{G}}(t^{(m)}=0) = \sum_{i} h_{\text{atomic}}(i)  \\
& h_{\text{atomic}}(i) \equiv
-\frac{1}{2}\left(\mu _e + \mu _g\right) n_{i} +\frac{U}{2} n_i^2  \\
&  \phantom{h_{\text{atomic}}(i) =}
+J \left\{ (T_i^x)^2 + (T_i^y)^2\right\} + J_z (T_i^z)^2  \\
& \phantom{h_{\text{atomic}}(i) =}
- \left(\mu _g -\mu _e\right) T^{z}_{i}  + U_{\text{diff}}  T^{z}_{i} n_{i}
\end{split}
\label{eqn:atmic-limit-Ham}
\end{equation}
with the following coupling constants
\begin{equation}
\begin{split}
& U=\frac{1}{4} (U_{\text{G}}^{(g)}+ U_{\text{G}}^{(e)} 
+2 V_{\text{G}}), \;\; 
U_{\text{diff}} = \frac{1}{2}(U_{\text{G}}^{(g)} - U_{\text{G}}^{(e)}) , \\
& J=V^{g\text{-}e}_{\text{ex}} , \;\;
J_{z} = \frac{1}{2}(U_{\text{G}}^{(e)}+ U_{\text{G}}^{(g)} -2 V_{\text{G}}), \\
& \mu_{m} = \frac{1}{2} (2 \mu_{\text{G}}^{(m)}+ U_{\text{G}}^{(m)}+V^{g\text{-}e}_{\text{ex}}) \quad 
(m=g,e) \; .
\end{split}
\label{eqn:Hund-by-Gorshkov}
\end{equation}

Let us consider the case of half-filling where each site is occupied by $N$ fermions on average. 
The Fermi statistics allows $(2N)!/(N!)^{2}$ states and, out of them, the optimal ones are chosen  
by the orbital-dependent terms [the last four terms in $h_{\text{atomic}}(i)$];  
when $N$ is even and $V^{g\text{-}e}_{\text{ex}}$ is positive, the states that transform under SU($N$) as 
the irreducible representation \eqref{eqn:Young-diagram-GS} are the ground states 
of $h_{\text{atomic}}(i)$\cite{Bois-C-L-M-T-15}.  
When $N=4$, they form the 20-dimensional representation of SU(4).  
For these states, the orbital pseudo-spin $\mathbf{T}_{i}$ is quenched and only the SU($N$) degree of 
freedom remains.  
When $N$ is odd, on the other hand, {\em both} SU($N$) spin and orbital are active and  
we obtain, in general, SU($N$)-orbital-coupled models.  In the following, we consider only the case 
with even-$N$ where pure spin models are obtained.    

Interactions among the remaining SU($N$) spins are derived by the second-order perturbation 
in $t^{(m)}$ as\cite{Bois-C-L-M-T-15}
\begin{equation}
\frac{1}{2} \left\{
\frac{{t^{(g)}}^{2}}{U+U_{\text{diff}}+J+\frac{J_z}{2}} 
+ \frac{{t^{(e)}}^{2}}{U-U_{\text{diff}}+J+\frac{J_z}{2}}
\right\} \mathcal{S}_i \cdot \mathcal{S}_{i+1}   \; ,
\label{eqn:2nd-order-effective-Ham-Gorshkov}
\end{equation}
where we have introduced a short-hand notation 
$\mathcal{S}_i \cdot \mathcal{S}_{i+1} \equiv \sum_{A=1}^{N^{2}-1} \mathcal{S}_{i}^{A}\mathcal{S}_{i+1}^{A}$ 
with $\mathcal{S}_{i}^{A}$ being the SU($N$) generators in the irreducible representation 
specified by the Young diagram \eqref{eqn:Young-diagram-GS}.  
Therefore, one sees that the model $\mathcal{H}_{\text{Heis}}$ [eq.\eqref{eqn:SUN-Heisenberg}] 
describes the low-energy physics of the alkaline-earth cold fermions [eq.\eqref{eqn:Gorshkov-Ham}] 
in the Mott-insulating phase (for $J=V^{g\text{-}e}_{\text{ex}}>0$).   
 
%%%%%%%%%%%%%%%%%%%%%%%%%%%%%%%%%%%%%%%%%%%%%%%%%%%%%%%%
\subsection{Solvable Hamiltonian}
\label{sec:solvable-Ham}
%%%%%%%%%%%%%%%%%%%%%%%%%%%%%%%%%%%%%%%%%%%%%%%%%%%%%%%%
Unfortunately, the Heisenberg Hamiltonian \eqref{eqn:SUN-Heisenberg} cannot be solved exactly. 
However, one can design a solvable model Hamiltonian whose ground state may share 
important properties with that of the original Heisenberg model \eqref{eqn:SUN-Heisenberg}.  
Clearly, when $N=2$, the Affleck-Kennedy-Lieb-Tasaki (AKLT) model proposed 
in Refs.~\onlinecite{Affleck-K-L-T-87,Affleck-K-L-T-88} will do the job: 
\begin{equation}
\mathcal{H}^{N=2}_{\text{VBS}} = \sum_i \left\{
  \mathbf{S}_i \cdot \mathbf{S}_{i+1}
+   \frac{1}{3} \left(\mathbf{S}_i \cdot \mathbf{S}_{i+1}\right)^2  \right\}  \; ,
\label{eqn:AKLT}
\end{equation}
where $\mathbf{S}_i$ denote the spin-1 operators.  
Its (rigorous) ground state, dubbed the valence-bond solid (VBS) state, 
is constructed\cite{Affleck-K-L-T-87,Affleck-K-L-T-88} by first decomposing an $S=1$ on each site 
into a pair of $S=1/2$s, forming uniform tiling of dimer singlets (`valence-bond solid') among the neighboring sites, 
and then fusing the $S=1/2$ pairs back to the original spin-1s.  
 
\begin{figure}[hbt]
\centering
\includegraphics[width=0.8\columnwidth,clip]{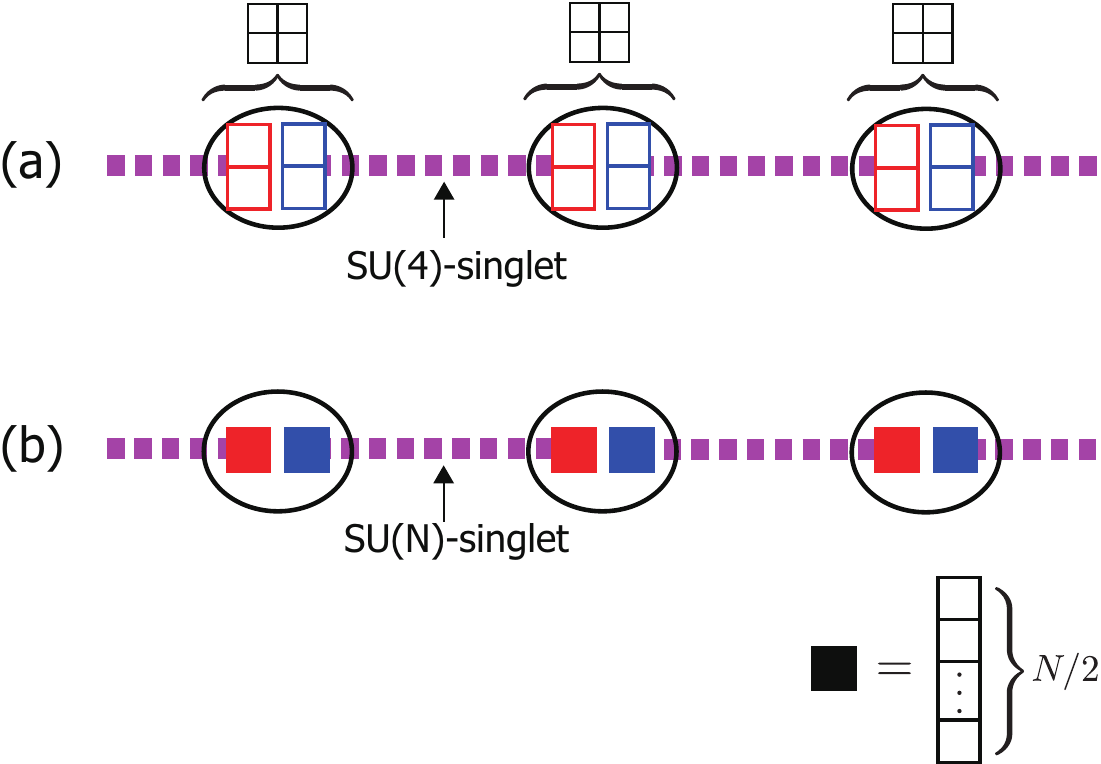}
\caption{(Color online) (a) Ground state of SU(4) VBS model [eq.\eqref{eqn:SU4-VBS}].  
Two 6-dimensional representations (`fictitious spins') are projected onto a physical 20-dimensional representations.  
(b) Similar construction applies to the cases with larger $N$ as well. 
\label{fig:SU(4)VBS}}
\end{figure}

Suggested by the above construction of the VBS ground state, we can think of constructing 
the model ground state by first preparing two auxiliary `spins' 
\begin{equation}
\text{\scriptsize $N/2$} \left\{ 
\yng(1,1,1)
 \right.  \quad (N=\text{even}) 
\label{eqn:Young-diagram-ancilla}
\end{equation}
on each site and pairing such spins on the adjacent sites into SU($N$) singlets (see Fig.~\ref{fig:SU(4)VBS}).  
The VBS ground state is obtained by projecting the product of the two fictitious spins on each site onto 
the physical Hilbert space characterized by the Young diagram in \eqref{eqn:Young-diagram-GS} 
(see Fig.~\ref{fig:SU(4)VBS}).  
In the following, we call this kind of states the SU($N$) VBS states.\footnote{%
In fact, there is another way of generalizing the spin-1 SU(2) VBS state.  
Instead of using two copies of the self-conjugate representations \eqref{eqn:Young-diagram-ancilla}, 
we may use the $n$-dimensional defining representation $\mathbf{n}$ and its conjugate $\bar{\mathbf{n}}$. 
This type of SU($N$) ``VBS state'' has been already discussed 
in the AKLT paper (Refs.~\onlinecite{Affleck-K-L-T-88}).}  
The parent Hamiltonians for these states read, e.g., for $N=4$\cite{Nonne-M-C-L-T-13,Bois-C-L-M-T-15} 
and for $N=6$ as\footnote{%
In fact, the expression of the parent Hamiltonian is {\em not} unique.  
There are 3 (6) free positive parameters in the parent Hamiltonian of the SU(4) [SU(6)] VBS state. 
The ones shown in the text are obtained when we require that they be of lowest degree in 
$\mathcal{S}{\cdot}\mathcal{S}$ and that the coefficient of the linear term be 1.}  
\begin{equation}
\begin{split}
& \mathcal{H}^{N=4}_{\text{VBS}} \\
&= \sum_i \left\{
  \mathcal{S}_i \cdot \mathcal{S}_{i+1}
+   \frac{13}{108} \left(\mathcal{S}_i \cdot \mathcal{S}_{i+1}\right)^2 
+ \frac{1}{216}\left(\mathcal{S}_i \cdot \mathcal{S}_{i+1}\right)^3  \right\} 
\end{split}
\label{eqn:SU4-VBS}
\end{equation}
and 
\begin{equation}
\begin{split}
\mathcal{H}^{N=6}_{\text{VBS}} =&  \sum_{i} \biggl\{
\mathcal{S}_i \cdot \mathcal{S}_{i+1} 
+ \frac{47}{508}(\mathcal{S}_i \cdot \mathcal{S}_{i+1} )^2  \\
& +\frac{17}{4572} (\mathcal{S}_i \cdot \mathcal{S}_{i+1} )^3 
+\frac{1}{18288} (\mathcal{S}_i \cdot \mathcal{S}_{i+1} )^4 
\biggr\} \; ,
\end{split}
\label{eqn:SU6-VBS}
\end{equation}
respectively.\footnote{%
For $N \ge 8$, the parent Hamiltonians are not always written {\em only} in terms of 
$\left(\mathcal{S}_i \cdot \mathcal{S}_{i+1}\right)$, as $\left(\mathcal{S}_i \cdot \mathcal{S}_{i+1}\right)$ alone 
cannot always distinguish among all the irreducible representations.}  
(In writing down the above expressions, we have normalized the generators 
$\mathcal{S}_{i}$ in such a way that the lengths of the simple roots are all $\sqrt{2}$.) 
The dimensions of the physical SU($N$) `spin' multiplet on each site are 
20 and 175 for $N=4$ and $6$, respectively.   
In Refs.~\onlinecite{Nonne-M-C-L-T-13,Bois-C-L-M-T-15}, the ground state wave function of 
$\mathcal{H}^{N=4}_{\text{VBS}}$ has been obtained in a matrix-product-state (MPS) form 
(see Appendix \ref{sec:MPS-matrices}).   
Clearly, the higher-order terms are rapidly suppressed as we go to larger-$N$.  
This suggests that the larger $N$ is, the better the VBS state shown in Fig.~\ref{fig:SU(4)VBS} 
approximates the ground state of the original Heisenberg model \eqref{eqn:SUN-Heisenberg}. 
This is quite natural in view of the large-$N$ results\cite{Read-S-NP-89,Read-S-90}. 
These models will serve as an ideal starting point for the study of the topological properties.  
 
%%%%%%%%%%%%%%%%%%%%%%%%%%%%%%%%%%%%%%%%%%%%%%%%%%
%%%%%%%%%%%%%%%%%%%%%%%%%%%%%%%%%%%%%%%%%%%%%%%%%%
\section{Symmetry-Protected Topological Phases}
\label{sec:SPT}
%%%%%%%%%%%%%%%%%%%%%%%%%%%%%%%%%%%%%%%%%%%%%%%%%%
In this section, we try to characterize the nature of the ground state of the SU($N$) spin chain 
\eqref{eqn:SUN-Heisenberg}.    
Specifically, in Sec.~\ref{sec:ES}, we show that the ground state of the model \eqref{eqn:SUN-Heisenberg} 
shares essentially the same properties with that of the solvable VBS models 
and that it is in fact in one of the SPT phases.   
Being topological, this class of topological phases defies the traditional characterization 
with broken symmetries and the associated local order parameters.  
One way is to use the {\em physical} edge states to distinguish between topological phases from trivial ones.  
However, this approach is not quite satisfactory in the following respects. First, even topologically {\em trivial} 
states may have certain structures around the edges of the system, as, e.g., the spin-2 Heisenberg 
chain does\cite{Nishiyama-T-H-S-95,Qin-N-S-95}.  
Second, in order to see the edge excitations, it is necessary 
to consider the excitation spectrum, while the topological properties are intrinsic to the ground state 
itself and should be seen only by examining the ground-state wave function. 

Recently, the use of the entanglement spectrum in characterizing topological phases 
has been suggested in Ref.~\onlinecite{Li-H-08}.  
This is based on the observation that the entanglement spectrum {\em resembles} 
the spectrum of the physical edge excitations. 
The idea has been successfully applied to various systems\cite{Pollmann-T-B-O-10,Pollmann-B-T-O-12,%
Fidkowski-K-11,Turner-P-B-11,Zheng-Z-X-L-11,Lou-T-K-K-11} and enabled us to characterize 
topological phases and quantum phase transitions among them. 
In this section, we present a clear evidence from the entanglement spectrum 
that the ground state of the SU(4) Heisenberg model 
\eqref{eqn:SUN-Heisenberg} is indeed in the SPT phases protected by SU(4) [PSU(4), precisely] symmetry. 
%%%%%%%%%%%%%%%%%%%%%%%%%%%%%%%%%%%%%%%%%%%%%%%%%%
\subsection{Haldane phase --an SPT primer}
%%%%%%%%%%%%%%%%%%%%%%%%%%%%%%%%%%%%%%%%%%%%%%%%%%
To understand the nature of the SPT phases in the case of SU($N$) symmetry, 
it is convenient to begin with the simplest case $N=2$.  
In 1983, Haldane conjectured\cite{Haldane-PLA-83,Haldane-PRL-83} that the ground-state properties of 
the spin-$S$ Heisenberg chain are qualitatively different according to the parity of $2S$; 
when $2S=\text{even}$, the ground state is in a featureless non-magnetic phase 
({\em Haldane phase}) with the gapped triplon excitations in the bulk,  
while, for odd $2S$, we have a gapless (i.e., algebraic) ground state with spinon excitations.   
This conjecture has been later confirmed both by the construction of 
a rigorous example\cite{ Affleck-K-L-T-87,Affleck-K-L-T-88,Arovas-A-H-88} [Eq.~\eqref{eqn:AKLT}] 
and by extensive numerical simulations\cite{White-H-93,Schollwock-G-J-96,Todo-K-01}.  
Soon after, it has been pointed out that the featureless gapped ground state of the integer-$S$ spin chains 
may have a {\em hidden} ``topological'' order characterized 
by non-local order parameters\cite{denNijs-R-89,Girvin-A-89,Kennedy-T-92-PRB,Kennedy-T-92-CMP} 
{\em at least} when $S$ is an odd integer\cite{Oshikawa-92}.  

However, it was not until the concept of SPT phases was established 
that the true meaning of ``topological order'' in the Haldane phase was understood\cite{Gu-W-09}.   
Now it is realized that the gapped phases in integer-spin chains with some protecting symmetry 
(e.g., time-reversal, reflection) are further categorized 
into topological phases and the other trivial ones.  
To understand the difference, it is useful to consider how the ground state in question transforms 
under the symmetry operation.  As the ground state is assumed symmetric, {\em the bulk} does not 
respond to the symmetry operation but the edges do.  As the consequence, the symmetry operation 
gets {\em fractionalized} into two pieces; one acts on the left edge and the other on the right. 
For instance, the VBS ground state $|S=1\text{ VBS}\rangle_{\alpha,\beta}$ of the spin-1 AKLT model \eqref{eqn:AKLT}  
hosts two {\em emergent} $S=\frac{1}{2}$ spins (i.e., $\alpha,\beta=\uparrow,\downarrow$) 
on both edges and hence transforms under the SO(3) rotation as 
\begin{equation}
|S=1\text{ VBS}\rangle_{\alpha,\beta} \xrightarrow{\text{SO(3)}} 
\sum_{\alpha^{\prime},\beta^{\prime}}U^{\dagger}_{\alpha,\alpha^{\prime}}U_{\beta,\beta^{\prime}}
|S=1\text{ VBS}\rangle_{\alpha^{\prime},\beta^{\prime}} \; ,
\end{equation}
where $U$ is the $S=\frac{1}{2}$ rotation matrix of SU(2).  
Putting it another way, $U$ serves as the mathematical labeling of the physical edge states. 
It is important to note that $U$ in the above in general is a projective representation 
of SO(3) as both $U^{\dagger}$ and $U$ appear simultaneously in the equation. 

Since this $U$ belongs to a non-trivial projective representation 
that is intrinsically different from any irreducible representations of the original SO(3), one sees 
that $|S=1\text{ VBS}\rangle_{\alpha,\beta}$ is in a non-trivial topological phase with emergent edge states.  
On the other hand, one can construct another exact ground state of a spin-1 chain which transforms 
as above but with $U$ belonging to the spin-1 representation.  Since the spin-1 representation is trivial 
in the sense of projective representation of SO(3), one can kill the would-be edge states by continuously 
deforming the Hamiltonian\cite{Pollmann-B-T-O-12} and this ground state is in a trivial phase.  
This reasoning may be readily generalized; when $U$ transforms like a half-odd-integer spin, 
the phase is topological, while when $U$ transforms in an integer-spin representation [i.e., linear representation 
of SO(3)], the system is in a trivial phase.   What is crucial in the topological properties is not the bulk 
spins at the individual sites but the {\em edge} spins.  

For later convenience, we summarize the situation in terms of Young diagrams.  
The spin-$S$ representation of SU(2) is represented by the following Young diagram:
\begin{equation}
\underbrace{%
\yng(2) \cdots \yng(1)
}_{\text{$2S$ boxes}} \; .
\end{equation}
With this in mind, the above result may be summarized as follows; when $U$ belongs to the representations 
\begin{equation}
\yng(1) \, , \; \yng(3) \, , \ldots \; , 
\end{equation}
the state represented by the corresponding MPS is topologically non-trivial, while the phase is 
trivial for $U$ transforming in 
\begin{equation}
\;\;
\yng(2)\, , \; \yng(4) \, , \ldots \; .
\end{equation}
That is, the number of boxes (mod 2) in the Young diagram for the representation 
to which $U$ belongs labels the topological classes protected by SO(3) and leads to 
the $\mathbb{Z}_{2}$ classification of the SO(3) SPT phases\cite{Chen-G-W-11}. 

%%%%%%%%%%%%%%%%%%%%%%%%%%%%%%%%%%%%%%%%%%%%%%%%%%
\subsection{SU($N$) topological phases}
\label{sec:SUN-SPT}
%%%%%%%%%%%%%%%%%%%%%%%%%%%%%%%%%%%%%%%%%%%%%%%%%%
Using the MPS representation\cite{Garcia-V-W-C-07} of the gapped ground state in one dimension, 
the above ``physical'' idea can be generalized and made mathematically precise.  
In fact, when a given ground state that is represented by an MPS
\begin{equation}
\sum_{\{m_{i}\}}A(m_1)A(m_2) \cdots A(m_L)|m_1\rangle{\otimes}\cdots {\otimes} |m_{L}\rangle
\end{equation}
is invariant under some symmetry $G$, 
a $D$-dimensional unitary matrix $U_{g}$ ($g \in G$) exists such that\cite{Garcia-W-S-V-C-08} 
\begin{equation}
A(m_i) \xrightarrow{G} \be^{i\phi_{g}} U^{\dagger}_{g} A(m_i) U_{g} \; ,
\label{eqn:MPS-UAU}
\end{equation}
where $A(m_i)$ denotes the $D{\times}D$ MPS matrices corresponding to the local physical state $|m_i\rangle$ and 
$\be^{i\phi_{G}}$ is a phase that depends on $G$.   
As has been mentioned above, the unitary matrix $U_{g}$ is in fact a projective representation 
of the symmetry $G$, that corresponds to the physical edge states\cite{Pollmann-T-B-O-10}.  
Therefore, the enumeration of topologically stable phases in the presence of symmetry $G$ boils down to 
counting the possible (non-trivial) projective representations of $G$.\cite{Chen-G-W-11}  

This problem was solved for SU($N$) and other Lie groups 
in Ref.~\onlinecite{Duivenvoorden-Q-13} and the picture 
in the previous section basically generalizes to the case of SU($N$) with some mathematical complications.  
Now the role of SO(3) in the previous section is played by 
$\text{PSU($N$)} \simeq \text{SU($N$)}/\mathbb{Z}_{N}$ [note $\text{SO(3)}\simeq \text{PSU(2)}$].  
Considering $\text{PSU($N$)}$ instead of SU($N$) amounts to restricting ourselves only to 
the irreducible representations of SU($N$) specified by Young diagrams with the number of boxes 
$n_{\text{Y}}$ divisible by $N$ [i.e., $n_{\text{Y}}=N k$ ($k=0,1,\ldots$)].   
This subset of irreducible representations roughly corresponds to the integer-spin ones in the SU(2) case.  
As in the previous section, 
the topological class of a given ground state (typically written as an MPS) is determined by 
looking at to which projective representation the unitary $U_{g}$ of the state belongs.  
Since inequivalent projective representations of PSU($N$) are labeled 
by $n_{\text{Y}}$ (mod $N$)\cite{Duivenvoorden-Q-13}, there are $N-1$ non-trivial topological 
classes (as well as one trivial one) specified by the $\mathbb{Z}_{N}$ label $n_{\text{top}}=n_{\text{Y}}$ (mod $N$).   
In the following, we use the name ``class-$n_{\text{top}}$'' for these topological classes 
(the class-0 corresponds to trivial phases).   
For instance, one can readily see that the ``VBS states'' (which are different from ours) 
investigated in Refs.~\onlinecite{Affleck-K-L-T-88,Katsura-H-K-08,Orus-T-11} fall into the class-1 and $N-1$ 
of the PSU($N$) SPT phases (see \href{}{Supplementary Material}).  
Quite recently, the class-1,2 phases as well as other (conventional) phases of SU(3)-invariant spin chains 
were investigated from the SPT point of view\cite{Morimoto-U-M-F-14}.   

A remark is in order about the definition of the topological class. 
In contrast to the SU(2) case where all the irreducible representations are self-conjugate, 
we must distinguish between an irreducible representation and its conjugate in SU($N$).  
The relation \eqref{eqn:MPS-UAU} suggests that if we have the edge state transforming under 
the projective representation $\mathcal{R}$ on the right edge, we necessarily have 
its conjugate $\bar{\mathcal{R}}$ on the other.   This means that when we talk about 
the topological class we must first fix which edge state we use to label the topological phases.  
Throughout this paper, 
we define the topological class by the {\em right edge state} [i.e., $U_{g}$ acting from the right in 
Eq.~\eqref{eqn:MPS-UAU}].  
Now it is easy to see that the SU($N$) VBS state introduced in Sec.~\ref{sec:solvable-Ham} 
belongs to class-$N/2$.  
%%%%%%%%%%%%%%%%%%%%%%%%%%%%%%%%%%%%%%%%%%%%%%%%%%
\subsection{Entanglement spectrum}
\label{sec:ES}
%%%%%%%%%%%%%%%%%%%%%%%%%%%%%%%%%%%%%%%%%%%%%%%%%%
Remarkably, the above-mentioned difference in the projective representation $U_g$ 
can be seen in the entanglement spectrum\cite{Pollmann-T-B-O-10}.  
In order to define the entanglement spectrum, we first divide the system 
into two subsystems A and B. Then, the entanglement spectrum $\{\xi_{\alpha}(\geq 0)\}$ 
is defined through the Schmidt decomposition of the ground state $|\psi\rangle$ of the entire system:
\begin{equation}
|\psi\rangle= \sum_{\alpha = 1}^{\chi} 
\be^{-\frac{\xi_{\alpha}}{2}} |\phi_{\alpha}^{\text{A}} \rangle \otimes |\phi_{\alpha}^{\text{B}} \rangle,
\end{equation}
where $\{|\phi_{\alpha}^{\text{A}} \rangle\}$ and $\{|\phi_{\alpha}^{\text{B}} \rangle\}$ are orthonormal 
basis sets for the subsystems satisfying 
$\langle \phi_{\alpha}^{\text{A,B}}|\phi_{\beta}^{\text{A,B}}\rangle = \delta_{\alpha\beta}$ and 
the number $\chi$ of finite $\xi_{\alpha}(< \infty)$ defines the Schmidt number. 

According to Ref.~\onlinecite{Li-H-08}, the entanglement spectrum of a given system exhibits 
a structure quite similar to that of the (energy) spectrum of the physical edge state of the same system and 
might be useful in characterizing topological states of matter. 
In one dimension, the edge states are not dispersive and we expect a discrete set of degenerate levels to appear 
in the entanglement spectrum reflecting the physical gapless edge modes. 
In fact, in accordance with the degeneracy in the entanglement spectrum, 
the projective representation $U_{g}$ assumes a block-diagonal structure\cite{Sanz-W-G-C-09}, where 
each block corresponds to an irreducible representation of SU($N$) compatible with the topological class.  
For instance, in a ground state in the class-2 topological phase of SU(4), each entanglement level 
should exhibit the degeneracy corresponding to an SU(4) irreducible representation with $n_{\text{Y}}=2$ (mod 4).  
In Table \ref{tab:Young}, the Young diagrams as well as their dimensions  
are listed for some typical irreducible representations compatible with the class-2 topological phase [i.e., 
$n_{\text{Y}}=2$ (mod 4)].  

%%%%%%%%%%%%%%%%%%%%%%%%%%%%%%%%%%%%%%%%%%%%%%%%%%
\subsubsection{VBS point}
%%%%%%%%%%%%%%%%%%%%%%%%%%%%%%%%%%%%%%%%%%%%%%%%%%
To investigate the topological phase protected by PSU(4) symmetry, we begin with the simplest case. 
The ground state of the SU(4) VBS Hamiltonian (\ref{eqn:SU4-VBS}) can be given exactly  
in the form of an MPS\cite{Nonne-M-C-L-T-13,Bois-C-L-M-T-15}   
and its entanglement spectrum is readily obtained by rendering the MPS into 
the canonical form (for the expressions of the matrices, see Appendix \ref{sec:MPS-matrices}).    

Reflecting the existence of the 6-dimensional (physical) edge states ${\tiny \yng(1,1)}$ ($n_{\text{top}}=n_{\text{Y}}=2$), 
the only entanglement level indeed is 6-fold degenerate indicating the class-2 
phase\cite{Nonne-M-C-L-T-13}: $\xi_{\alpha}=\log 6$ ($\alpha=1,\ldots,6$; $\chi=6$).  
This is in perfect agreement with the above argument.  

%%%%%%%%%%%%%%%%%%%%%%%%%%%%%%%%%%%%%%%%%%%%%%%%%%
\subsubsection{Heisenberg point}
%%%%%%%%%%%%%%%%%%%%%%%%%%%%%%%%%%%%%%%%%%%%%%%%%%
In order to check if the ground state of the SU(4) Heisenberg chain \eqref{eqn:SUN-Heisenberg} 
is in the class-2 topological phase, we calculated  
the entanglement spectrum with the infinite time-evolving block decimation (iTEBD) 
algorithm\cite{Vidal-iTEBD-07,Orus-V-08}.   
which enables us to directly access the entanglement spectrum. 

\begin{figure}[hbt]
\centering
\includegraphics[width=0.9\columnwidth,clip]{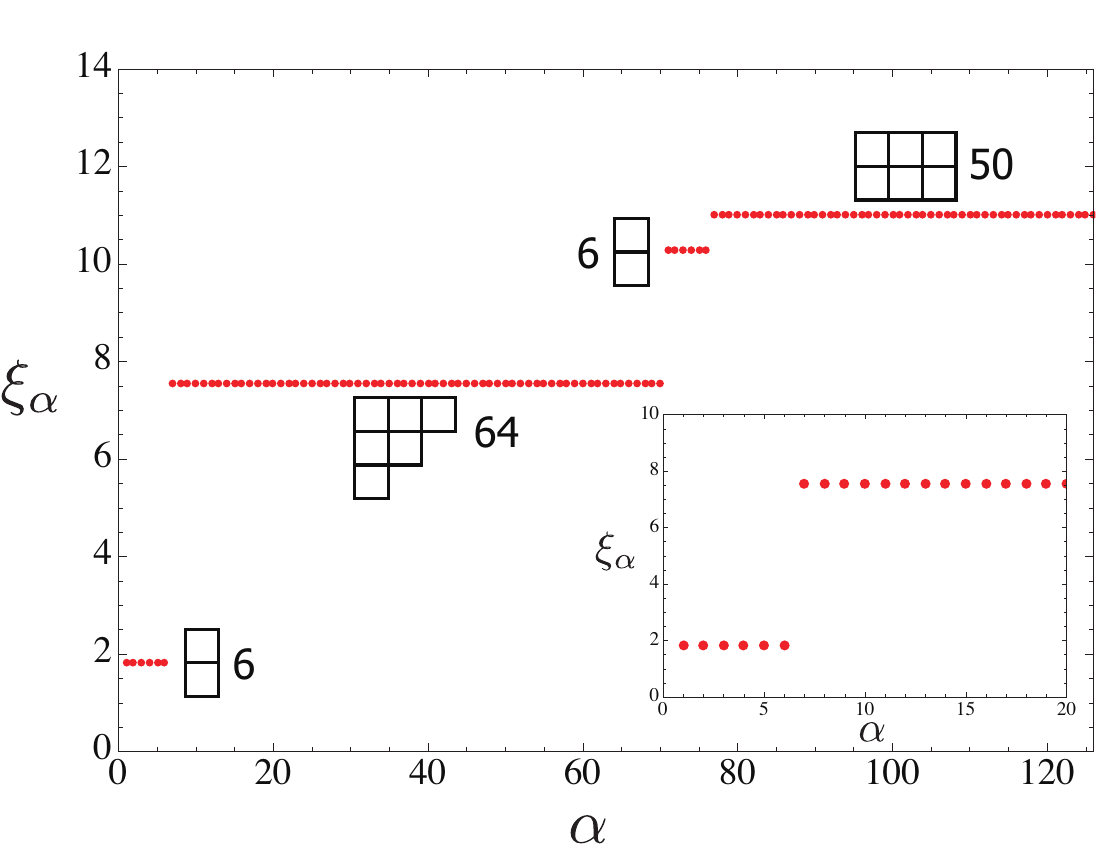}
\caption{(Color online) Entanglement spectrum of an infinite SU(4) Heisenberg chain 
calculated by iTEBD.  The degeneracy $\{6, 64, 6, 50\}$ may be understood in terms 
of the SU(4) irreducible representations shown in the figure. 
(inset) Zoom-up of the lowest six-fold-degenerate entanglement level. 
\label{fig:ES_a=00}}
\end{figure}

The simulations were done using the MPS with the bond dimensions up to 150 and 
the spectrum obtained is shown in Fig.~\ref{fig:ES_a=00}.  
The degrees of degeneracy seen in Fig.~\ref{fig:ES_a=00} are $\{6, 64, 6, 50\}$ from 
the bottom to the top.  
Clearly, this pattern perfectly fits into the dimensions in Table \ref{tab:Young}; the edge state transform 
under the four (self-conjugate) irreducible representations shown in Fig.~\ref{fig:ES_a=00}.  
All these have $n_{\text{Y}}=2$ (mod $4$) and, 
from the discussion in Sec.~\ref{sec:SUN-SPT}, this ground state is classified as the topological class 2.

Here a remark is in order.  As the bosonic SU(4) Heisenberg model \eqref{eqn:SUN-Heisenberg} is obtained 
as the effective Hamiltonian in the Mott phase of the {\em fermionic} model \eqref{eqn:Gorshkov-Ham}, 
one may suspect that the same degeneracy structure could have been obtained for the original fermion model as well.  
However, this is not necessarily the case.  In fact, in models where both bosonic and fermionic 
modes coexist, the entanglement spectrum contains the contribution from the fermionic sector 
as well as that from the bosonic one, and some of the levels may not obey 
the degeneracy rule that is obtained for the {\em purely} bosonic models\cite{Hasebe-T-13}.  
This is the reason why we simulated the effective bosonic model \eqref{eqn:2nd-order-effective-Ham-Gorshkov}.  
%%%%%%%%%%%%%%%%%%%% TABLE 1 %%%%%%%%%%%%%%%%%%%%%%%%%%%%%%%%
\begin{table}[htb]
\caption{\label{tab:Young} Typical Young diagrams with the number of boxes $n_{\text{Y}}\equiv 2$ (mod 4) 
 and their dimensions in SU(4).}
\begin{ruledtabular}
\begin{tabular}{cccc}
$n_{\text{Y}}$ & Young diagram & dimension  \\
\hline
2 & 
\multicolumn{3}{c}{%
\begin{tabular}{@{}lp{18mm}r@{}}
{\tiny \yng(1,1)} && 6 \\
{\tiny \yng(2)}  && 10
\end{tabular} 
} \\
\hline
6 & 
\multicolumn{3}{@{}c@{}}{%
\begin{tabular}{@{}lp{10mm}r@{}}
{\tiny \yng(2,2,2)} &   & 10 \\
{\tiny \yng(3,3)} & & 50 \\
{\tiny \yng(3,2,1)} & & 64 \\
{\tiny \yng(4,1,1)} & & 70 \\
{\tiny \yng(6)} & & 84 \\
{\tiny \yng(4,2)} & & 126 \\
{\tiny \yng(5,1)} & & 140
\end{tabular} 
} 
\\
\hline
10 & 
\multicolumn{3}{@{}c@{} }{%
\begin{tabular}{@{}lp{5mm}r@{}}
{\tiny \yng(4,3,3)} & & 70 \\
{\tiny \yng(4,4,2)} &  &126 \\
{\tiny \yng(5,5)} & & 196 \\
{\tiny \yng(6,2,2)} & &  270 \\
{\tiny \yng(10)} & & 286 
\end{tabular}
} \\
\end{tabular}
\end{ruledtabular}
\end{table}
%%%%%%%%%%%%%%%%%%%%%%%%%%%%%%%%%%%%%%%%%%%%%%%%%%%%%%%%%
%%%%%%%%%%%%%%%%%%%%%%%%%%%%%%%%%%%%%%%%%%%%%%%%%%%%%
%%%%%%%%%%%%%%%%%%%%%%%%%%%%%%%%%%%%%%%%%%%%%%%%%%%%%
%%%%%%%%%%%%%%%%%%%%%%%%%%%%%%%%%%%%%%%%%%%%%%%%%%
\subsubsection{Continuity between Heisenberg and VBS points}
\begin{figure}[htb]
\centering
\includegraphics[width=1.0\columnwidth,clip]{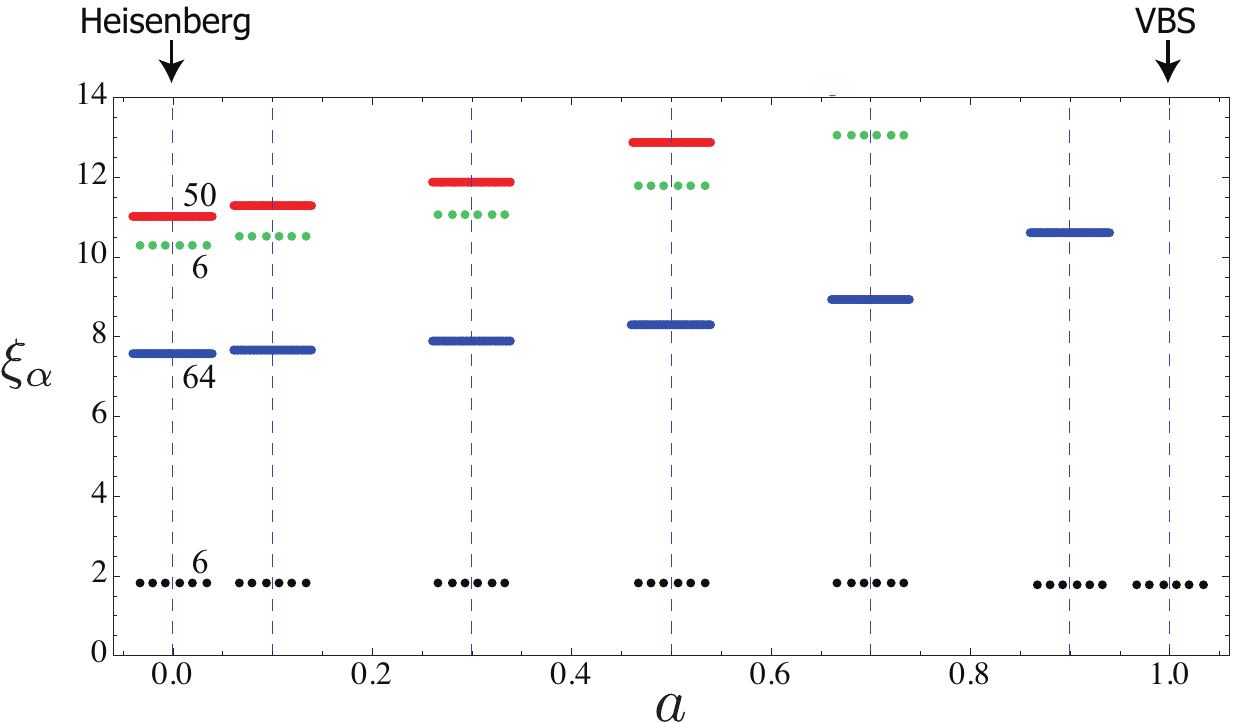}
\caption{(Color online) Evolution of entanglement spectrum as we interpolate between 
SU(4) Heisenberg model [\eqref{eqn:SUN-Heisenberg}; $a=0$] and SU(4) VBS model 
[\eqref{eqn:SU4-VBS}; $a=1$].  Numbers shown next to the levels are degrees of degeneracy.  
\label{fig:SU4-ES-all}}
\end{figure}

In the previous sections, we have seen, by inspecting the entanglement spectra, that 
the original SU(4) Heisenberg model \eqref{eqn:SUN-Heisenberg} and the solvable SU(4) VBS model 
\eqref{eqn:SU4-VBS} share the same topological properties in common. 
Next, we consider adiabatic connection between the Heisenberg point and the solvable VBS point 
to show that they belong to the same unique phase in the sense that they are connected to each other 
without quantum phase transitions\cite{Chen-G-W-10}. 
To connect the two Hamiltonians, we use the following one-parameter family of Hamiltonians
\begin{equation}
\begin{split}
& \mathcal{H}(a) \\
& = \sum_i \left\{%
 \mathcal{S}_i {\cdot} \mathcal{S}_{i+1}  
+a\left[   \frac{13}{108} \left(\mathcal{S}_i {\cdot} \mathcal{S}_{i+1}\right)^2 
+ \frac{1}{216}\left(\mathcal{S}_i {\cdot} \mathcal{S}_{i+1}\right)^3  \right] 
\right \} \; ,
\end{split}
\label{eqn:Ha}
\end{equation}
where $a$ is an interpolating parameter changing from 0 [Heisenberg point: Eq. \eqref{eqn:SUN-Heisenberg}] 
to 1 [VBS point: Eq.\eqref{eqn:SU4-VBS}].  
We calculated the entanglement spectrum of the ground state of $\mathcal{H}(a)$ 
for $a=0.0$, $0.1$, $0.3$, $0.5$, $0.7$, $0.9$, and $1.0$ 
with iTEBD and the results are shown in Fig.~\ref{fig:SU4-ES-all}.   
It is evident that the structure of the entanglement spectrum (including the six-fold degeneracy 
in the lowest level) is preserved all the way from the Heisenberg point up to the VBS point showing 
that the two models indeed belong to the same class-2 topological phase.  
%%%%%%%%%%%%%%%%%%%%%%%%%%%%%%%%%%%%%%%%%%%%%%%%%%
%%%%%%%%%%%%%%%%%%%%%%%%%%%%%%%%%%%%%%%%%%%%%%%%%%%%%
%%%%%%%%%%%%%%%%%%%%%%%%%%%%%%%%%%%%%%%%%%%%%%%%%%%%%

%%%%%%%%%%%%%%%%%%%%%%%%%%%%%%%%%%%%%%%%%%%%%%%%%%
\section{Non-local string order parameters}
\label{sec:string-orer-parameter}
%%%%%%%%%%%%%%%%%%%%%%%%%%%%%%%%%%%%%%%%%%%%%%%%%%
In Sec.~\ref{sec:ES}, we have seen that the structure of the entanglement spectrum 
helps us to identify the topological class of a given ground state provided that we have enough 
information on the protecting symmetry of the system in advance. 
However, in general, the degeneracy structure alone does {\em not} uniquely identify the topological class.   
For instance, the class-2 phase of PSU(4)-symmetric systems has doubly-degenerate 
entanglement levels (see Appendix \ref{sec:PSUN-to-ZnxZn}), 
that are reminiscent of the Haldane phase protected by 
$\mathbb{Z}_{2}{\times}\mathbb{Z}_{2}$, although these two phases are essentially different 
as we will see in Sec.~\ref{sec:symmetry-reduction}.  
Furthermore, despite some recent 
proposals\cite{Guhne-H-B-E-L-M-S-02,Abanin-D-12,Daley-P-S-Z-12,Pichler-B-D-L-Z-13},  
it is not very straightforward to directly measure entanglement in experiments.  
In fact, what is more fundamental in identifying SPT phases is the projective representation $U_{g}$. 
Therefore, ``order parameters'' that have more direct access to $U_{g}$ is desirable.   

Several order parameters for SPT phases, including a gauge-invariant product of $U_{g}$s, 
were proposed recently\cite{Pollmann-T-12} (for discussion of the detection of SPTs using 
the response of the physical edge states to external perturbations, see Ref.~\onlinecite{Liu-C-W-11}). 
However, these order parameters are written directly in terms of the projective representation 
$U_{g}$ and are not accessible in experiments in spite of their use in numerical simulations.  
Therefore, for the purpose of the detection of SPT phases in experiments, the characterization with 
order parameters, that are written in terms of {\em measurable} quantities, is still useful.   
In this section, we introduce a set of non-local string order parameters for our SU($N$) spin system  
to characterize the topological phases.   
%%%%%%%%%%%%%%%%%%%%%%%%%%%%%%%%%%%%%%%%%%%%%%%%%%
\subsection{$\mathbb{Z}_{N}{\times}\mathbb{Z}_{N}$ and SPT phases}
\label{sec:def-ZnxZn}
%%%%%%%%%%%%%%%%%%%%%%%%%%%%%%%%%%%%%%%%%%%%%%%%%%
In Ref.~\onlinecite{Duivenvoorden-Q-ZnxZn-13}, a set of generalized string order parameters 
based on the symmetry $\mathbb{Z}_{N}{\times}\mathbb{Z}_{N}$ was introduced for 
generic $\mathbb{Z}_{N}{\times}\mathbb{Z}_{N}$-invariant systems 
and its connection to the $\mathbb{Z}_{N}{\times}\mathbb{Z}_{N}$ SPT phases was discussed.  
As PSU($N$) and $\mathbb{Z}_{N}{\times}\mathbb{Z}_{N}$ have  
the same cohomology group\cite{Duivenvoorden-Q-ZnxZn-13,Chen-G-L-W-13} 
$H^{2}(\text{PSU($N$)},\text{U(1)})=H^{2}(\mathbb{Z}_{N}{\times}\mathbb{Z}_{N},\text{U(1)})=\mathbb{Z}_{N}$ 
in common, 
we may expect that we can characterize our topological phase by using these string order parameters.  
In order to adapt the string order parameters, that was introduced in Ref.~\onlinecite{Duivenvoorden-Q-ZnxZn-13} 
in the context of $\mathbb{Z}_{N}{\times}\mathbb{Z}_{N}$-invariant systems,  
to our SU($N$) case, we have to first identify the two {\em commuting} $\mathbb{Z}_{N}$s in 
SU($N$).   
%\textcolor[rgb]{1,0,0}{%
%Since the final results depend on $N$, we focus on the case of SU(4) in the following. }%%

The construction of a pair of $\mathbb{Z}_{N}$s itself does not rely on a particular choice 
of the irreducible representation.   
In fact, we do not need the explicit expressions of the generators which depend on 
the choice of the basis and representation; the commutation relations among 
the generators suffice for our purpose.  
The most convenient way is to use the Cartan-Weyl basis $\{ H_{a},E_{\alpha}\}$ 
that satisfy\cite{Georgi-book-99}
\begin{equation}
\begin{split}
&  [H_{a},H_{b}]=0 \, , \;\; 
[H_{a},E_{\alpha}] = (\alpha)_{a}E_{\alpha} \, , \\
& [E_{\alpha},E_{-\alpha}] = \sum_{a=1}^{3} (\alpha)_{a}H_{a} \, ,  \; 
\text{Tr}\, (H_{a}H_{b})= \kappa \delta_{ab} ,
\\
&(a,b=1,\ldots,N-1) 
\end{split}
\label{eqn:Cartan-Weyl-commutation-rel}
\end{equation}
where $\alpha$  denotes the $N^{2}-N$ roots of SU($N$) normalized as $|\alpha|=\sqrt{2}$ which are generated 
by the simple roots $\alpha_{i}$ ($i=1,2,3$).  
The normalization $\kappa$ depends on the representation and set to 1 for the $N$-dimensional fundamental 
representation ${\tiny \yng(1)}$ [e.g., $\kappa=16$ for the 20-dimensional representation ${\tiny \yng(2,2)}$ of SU(4) 
considered here].  
In the actual calculations, one may use, e.g.,  
the generators and the weights given in Sec.~13.1 of Ref.~\onlinecite{Georgi-book-99} 
with due modification of the normalization. 

Now let us look for the operators $G_{Q}$ and $G_{P}$ that generate the two $\mathbb{Z}_{N}$s.  
Regardless of $N$, the first generator $G_{Q}$, which is diagonal and plays the role of $S^{z}$ in SU(2),  
is given simply by 
\begin{equation}
G_{Q} = \sum_{k=1}^{N-1} (\vec{\rho})_{k} H_{k} \; ,
\label{eqn:def-G_Q}
\end{equation}
where $H_{k}$ are the $N-1$ Cartan generators and 
$\vec{\rho}$ is the Weyl vector of SU($N$).  The generator $G_{Q}$ 
has the following simple commutation relations with the simple roots $\alpha$:
\begin{equation}
[ G_{Q}, E_{\pm \alpha}]= \pm E_{\alpha} \; ,
\label{eqn:comm-rel-Gq-E}
\end{equation}
which guarantee integer-spaced eigenvalues of $G_{Q}$ 
(for the fundamental representation $\boldsymbol{N}$, they are essentially $1,2,\cdots, N$).  
With this, the first $\mathbb{Z}_{N}$ is generated as
\begin{equation}
Q = c_{N} \exp \left(i\frac{2\pi}{N} G_{Q}\right) \; ,
\end{equation}
where the phase $c_{N}$ has been introduced so that $Q$ satisfy $Q^{N}=1$.  
The expression of the other generator $G_{P}$ depends on $N$ and, in the following, we will explicitly 
work it out for $N=4$.  

The first $\mathbb{Z}_{4}$-generator $Q$ is defined in terms of the two commuting SU(4) 
generators (the Cartan generators) as
\begin{equation}
\begin{split}
& Q \equiv \be^{i\frac{3\pi}{4}} \exp\left( i \frac{2\pi}{4}G_{Q} \right)  , \;\; Q^{4}=1 \\
& G_{Q} \equiv  2H_{1} + H_{2}  \; .
\end{split}
\label{eqn:Q-by-SU4}
\end{equation}
The generator $G_{Q}$ satisfies Eq.~\eqref{eqn:comm-rel-Gq-E}.  
On the other hand, the second $\mathbb{Z}_{4}$ is generated by
\begin{equation}
\begin{split}
& P \equiv \be^{i\frac{3\pi}{4}} \exp\left( i \frac{2\pi}{4}G_{P} \right)  , \;\; P^{4}=1 \\
& G_{P} \equiv  - \frac{1}{2} \sum_{\alpha} E_{\alpha}  + \frac{i}{2} \left( \sum_{i=1}^{3}E_{\alpha_{i}}  
- E_{\alpha_{1}+\alpha_{2}+\alpha_{3}} \right)   \\
& \phantom{G_{P} \equiv}
- \frac{i}{2} \left( \sum_{i=1}^{3}E_{-\alpha_{i}}- E_{-\alpha_{1}-\alpha_{2}-\alpha_{3}} \right)   \; .
\end{split}
\label{eqn:P-by-SU4}
\end{equation}
The summation $\sum_{\alpha}$ runs over all the twelve non-zero roots $\alpha$ of SU(4).  
Here we do not give the explicit expressions of the generators which depend on 
a particular choice of the basis, since giving the commutation relations \eqref{eqn:Cartan-Weyl-commutation-rel} 
suffices to define $\mathbb{Z}_{4}{\times}\mathbb{Z}_{4}$ 
(see \href{}{Supplementary Material} for the expressions in a particular basis set 
that are more convenient for the actual calculations).  
It is important to note that the two operators $Q$ and $P$ constructed here 
generate $\mathbb{Z}_{4}{\times}\mathbb{Z}_{4}$ (i.e., $[Q,P]=0$) {\em only} when the number of 
boxes in the Young diagram is an integer multiple of 4.  
In other words, what we have defined is the $\mathbb{Z}_{4}{\times}\mathbb{Z}_{4}$ subgroup of PSU(4).   
This is reminiscent of that the two $\pi$-rotations 
along the $x$ and $z$ axes generate $\mathbb{Z}_{2}{\times}\mathbb{Z}_{2}$ only 
for $\text{SO(3)}\simeq \text{SU(2)}/\mathbb{Z}_{2}$.  
In Appendix \ref{sec:ZnxZn}, we present the expressions of $G_{P}$ and $G_{Q}$ for other $N$s.   

Having explicitly constructed a $\mathbb{Z}_{N}{\times}\mathbb{Z}_{N}$ subgroup of PSU($N$), 
we now consider how the existence of this subgroup leads to $(N-1)$ SPT phases.  
Consider the two $\mathbb{Z}_{N}$ generators 
$P$ and $Q$ satisfying 
\begin{equation}
(Q)^{N}=(P)^{N}=1 \; .
\label{eqn:U-to-N-1}
\end{equation}
As has been shown above, we can explicitly construct $P$ and $Q$ using 
the generators of SU($N$).  
By carefully choosing the gauge, we can always make the corresponding projective representations 
$U_{P}$ and $U_{Q}$ satisfy 
\begin{equation}
\begin{split}
& U_{1}=\mathbf{1} \, ,  \; (U_{P})^{N}= (U_{Q})^{N}=\mathbf{1} \, , 
\\
& U_{P^{n}}=(U_{P})^{n} \, , \; U_{Q^{n}}=(U_{Q})^{n}  \quad (n=1,\ldots, N-1) \; .
\end{split}
\label{eqn:ZnxZn-gauge-choice}
\end{equation}
If one requires that $QP=PQ$ hold when both sides act on the MPS in question, 
one obtains from Eq.~\eqref{eqn:MPS-UAU}
\begin{equation}
A(m) = (U_{Q}U_{P}U_{Q}^{\dagger}U_{P}^{\dagger})A(m) 
(U_{P}U_{Q}U_{P}^{\dagger}U_{Q}^{\dagger})   \; .
\end{equation}
When the MPS in question is pure and canonical, 
this implies 
\begin{equation}
%U_{P}U_{Q}U_{P}^{\dagger}U_{Q}^{\dagger} = \be^{i\Phi_{12}} \mathbf{1}  \;\; 
%\leftrightharpoons \;\; 
U_{P}U_{Q} =  \be^{- i\Phi_{QP}} U_{Q}U_{P} \; .
\label{eqn:U1-U2-exchange-Zn}
\end{equation}
On the other hand, combining $(U_{P})^{N}=1$ and 
$U_{P}U_{Q}U_{P}^{\dagger}U_{Q}^{\dagger} = \be^{ - i\Phi_{QP}} \mathbf{1}$ obtained above, 
we obtain another relation:
\begin{equation}
\begin{split}
U_{Q}U_{P}^{N-1} &=  U_{Q}U_{P}^{\dagger}
= U_{P}^{N-1}(U_{P}U_{Q}U_{P}^{\dagger}U_{Q}^{\dagger})U_{Q} 
 \\
& = \be^{ - i\Phi_{QP}}U_{P}^{N-1}U_{Q} \; .
\end{split}
\end{equation}
Using \eqref{eqn:U1-U2-exchange-Zn}, the right-hand side may be rewritten as
\begin{equation}
\begin{split}
\be^{ -i\Phi_{QP}}U_{P}^{N-1}U_{Q} &= (\be^{-i\Phi_{QP}})^{2}U_{P}^{N-2}U_{Q}U_{P}  \\
&= (\be^{-i\Phi_{QP}})^{N}U_{Q}U_{P}^{N-1}  \; .
\end{split}
\end{equation} 
Therefore, we arrive at the conclusion that $\be^{i\Phi_{QP}}$ is the $\mathbb{Z}_{N}$ 
phase\cite{Duivenvoorden-Q-ZnxZn-13}: 
%\textcolor[rgb]{1,0,0}{\bf Show the phase $\Phi_{PQ}$ here is equivalent to $n_{\text{top}}$.}
\begin{equation}
\be^{i\Phi_{QP}}= \be^{i \frac{2\pi}{N}n_{\text{top}}} 
= \omega^{n_{\text{top}}} \;\; (n_{\text{top}}=0,1,\ldots, N-1) \; .
\label{eqn:exchange-phase}
\end{equation}
To see that $\Phi_{QP}$ is in fact given by $(2\pi/N) n_{\text{top}}[=(2\pi/N) n_{\text{Y}}]$, we just note that 
$U_{Q}U_{P}=\be^{i\frac{2\pi}{N}} U_{P}U_{Q}$ for the $N$-dimensional representation ${\tiny \yng(1)}$ 
and that other representations are constructed by tensoring ${\tiny \yng(1)}$ $n_{\text{Y}}$ times.  
Eq.~\eqref{eqn:exchange-phase} implies that the exchange phase between $U_{P}$ and $U_{Q}$ carries the information 
on the topological class $n_{\text{top}}$.   
%%%%%%%%%%%%%%%%%%%%%%%%%%%%%%%%%%%%%%%%%%%%%%%%%%
\subsection{Definition}
%%%%%%%%%%%%%%%%%%%%%%%%%%%%%%%%%%%%%%%%%%%%%%%%%%
Next, we define another set of operators $\hat{X}_{P}$ and $\hat{X}_{Q}$ satisfying 
the following commutation relations with $\hat{P}$ and $\hat{Q}$ introduced in the previous section
\begin{equation}
\begin{split}
& \hat{Q}^{\dagger}\hat{X}_{Q} \hat{Q} = \omega \hat{X}_{Q} \;\; , \quad 
\hat{P}^{\dagger}\hat{X}_{Q} \hat{P} = \hat{X}_{Q} \\
& \hat{Q}^{\dagger}\hat{X}_{P} \hat{Q} = \hat{X}_{P} \;\; , \quad 
\hat{P}^{\dagger}\hat{X}_{P} \hat{P} = \omega^{-1}\hat{X}_{P}  
\quad 
(\omega=\be^{i\frac{2\pi}{N}})
\end{split}
\label{eqn:Z4xZ4-XpXq}
\end{equation}
for {\em any} irreducible representations of SU($N$).  
Using the commutation relations \eqref{eqn:Cartan-Weyl-commutation-rel}, one sees that 
the operators $\hat{X}_{Q}$ and $\hat{X}_{P}$ for $N=4$ can be expressed by the SU(4) 
generators as
\begin{equation}
\begin{split}
& \hat{X}_{Q} = \frac{1}{\sqrt{2}}( E_{-\alpha_1}+E_{-\alpha_2}+E_{-\alpha_3} + E_{-\alpha_4})   \\
& \hat{X}_{P} = H_{1} - i H_{3}    \; ,
\end{split}
\label{eqn:XQ-XP-by-SU4}
\end{equation}
where $\alpha_4 \equiv - \alpha_1 -\alpha_2 -\alpha_3$ and 
the normalization has been chosen such that 
$\text{Tr}\,({\hat{X}_{Q}}^{\dagger}\hat{X}_{Q})= \text{Tr}\,({\hat{X}_{P}}^{\dagger}\hat{X}_{P})$.  
From these operators, we define the following string operators:
\begin{subequations}
\begin{align}
V_{P}(m,n;i) & \equiv \hat{Q}^{\dagger}(1)^{n} \cdots \hat{Q}^{\dagger}(i-1)^{n} 
\left(\hat{X}_{P}(i)\right)^{m}  
\label{eqn:string-op-1} \\
V_{Q}(m,n;i) &\equiv \left(\hat{X}_{Q}(i)\right)^{m} \hat{P}(i+1)^{n} \cdots \hat{P}(L)^{n}   \; .
\label{eqn:string-op-2} 
\end{align}
\end{subequations}
Then, the string-order parameters (SOP) are (infinite-distance limits of) the two-point functions of these 
string operators:
\begin{subequations}
\begin{align}
\begin{split}
& \mathcal{O}_{1}(m,n) \equiv 
\lim_{|i-j|\nearrow \infty}  \langle V_{P}(m,n;i)V_{P}^{\dagger}(m,n;j) \rangle \\
& = \lim_{|i-j|\nearrow \infty} 
\Biggl\langle \left\{\hat{X}_{P}(i)\right\}^{m} \left\{
\prod_{i\leq k <j} \hat{Q}(k)^{n}
\right\} \left\{ \hat{X}_{P}^{\dagger}(j) \right\}^{m}  \Biggr\rangle
\label{eqn:def-stringOP-1b} 
\end{split}
\\
\begin{split}
& \mathcal{O}_{2}(m,n) \equiv \lim_{|i-j|\nearrow \infty} \langle V_{Q}(m,n;i)V_{Q}^{\dagger}(m,n;j) \rangle\\
& = \lim_{|i-j|\nearrow \infty} 
\Biggl\langle
\left\{  \hat{X}_{Q}(i) \right\}^{m}  
\left\{ \prod_{i < k \leq j} \hat{P}(k)^{n}
\right\} \left\{\hat{X}_{Q}^{\dagger}(j)  \right\}^{m} \Biggr\rangle  \\
& \qquad (0 \leq m,n < N) 
 \; . 
 \end{split}
\label{eqn:def-stringOP-2b}
\end{align}
\end{subequations}
The subscripts 1 and 2 refer to the SOP corresponding to the two commuting 
$\mathbb{Z}_{N}$'s (associated with $Q$ and $P$, respectively).  

It is important to note that when the model is realized in the cold-atom system \eqref{eqn:Gorshkov-Ham}, 
the SOP $\mathcal{O}_{1}(m,n)$ are expressed {\em only} in terms of the local fermion numbers 
$n_{\alpha,i}=c^{\dagger}_{g\alpha,i}c_{g\alpha,i}+c^{\dagger}_{e\alpha,i}c_{e\alpha,i}$.  
In fact, the expressions \eqref{eqn:def-stringOP-1b} involves only the diagonal 
generators $\{ H_{a} \}$ [see, e.g., Eqs.~\eqref{eqn:Q-by-SU4} and \eqref{eqn:XQ-XP-by-SU4}] 
which, when second-quantized, can be written only with the local fermion densities $n_{\alpha,i}$.  
This property is desirable in view of future detection of the non-local order 
with the site-resolved-imaging techniques.\cite{Endres-etal-stringOP-11,Gross-B-review-15}

As is seen in \eqref{eqn:def-stringOP-2b}, the second SOP $\mathcal{O}_{2}(m,n)$ contain 
the off-diagonal generators [see Eqs.~Eqs.~\eqref{eqn:P-by-SU4} and \eqref{eqn:XQ-XP-by-SU4}] and 
are more complicated; in order to express them 
in terms of the fermions, we first second-quantize the (off-diagonal) generators, e.g., as
\[  \hat{E}_{\alpha,i} = c^{\dagger}_{g\beta,i}(\mathcal{E}_{\alpha})_{\beta\gamma}c_{g\gamma,i}
+ c^{\dagger}_{e\beta,i}(\mathcal{E}_{\alpha})_{\beta\gamma}c_{e\gamma,i} \; ,\]
where the $4{\times}4$ matrices $\mathcal{E}_{\alpha}$ are four-dimensional fundamental representations 
of the generators $E_{\alpha}$ 
(see \href{}{Supplementary Material} for the expressions).  
Acting on the states in \eqref{eqn:Young-diagram-GS}, 
the second-quantized generators $\hat{E}_{\alpha}$ reproduce the ones appearing in \eqref{eqn:def-stringOP-2b}.  

\begin{figure}[htb]
\centering
\includegraphics[width=0.9\columnwidth,clip]{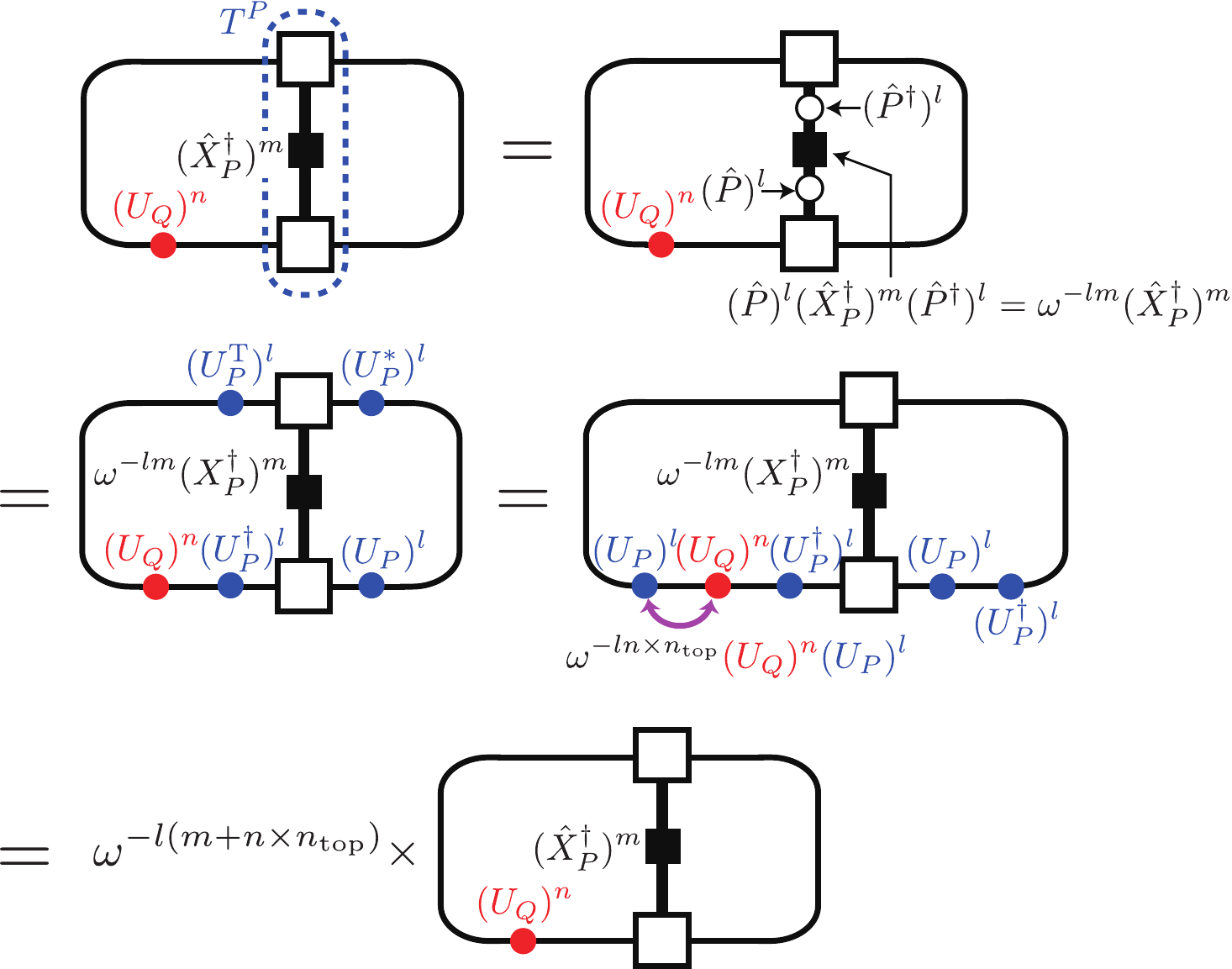}
\caption{(Color online) Boundary term $\mathcal{O}_{1,\text{R}}(m,n)$ 
carries the information on the exchange phase 
$\omega^{n_{\text{top}}}$ between $U_{P}$ and $U_{Q}$ [see Eq.~\eqref{eqn:U1-U2-exchange-Zn}].  
Here a trivial identity 
$(\hat{X}_{P})^{m}=(\hat{P}^{\dagger})^{l}\left\{ (\hat{P})^{l}(\hat{X}_{P})^{m}(\hat{P}^{\dagger})^{l}\right\}
(\hat{P})^{l}$ ($l$: arbitrary) has been used. 
\label{fig:string-boundary-SUN}}
\end{figure}

The merit of using the SOP is that they carry the information on the projective representation 
$U_{P}$ and $U_{Q}$ that determine the topological class\cite{Pollmann-T-12,Hasebe-T-13} 
(see Sec.~\ref{sec:def-ZnxZn}).  
To show this, we first note that the SOP decouple into the product 
of the boundary contributions:
\begin{equation}
\begin{split}
& \mathcal{O}_{1}(m,n) \xrightarrow{|i-j|\nearrow \infty}
\sum_{\alpha,\beta}\left\{
(T_{Q}^{X_P} \mathbf{V}^{(Q)}_{\text{R},1})
(\mathbf{V}^{(Q)}_{\text{L},1}T^{X_P})
\right\}_{\alpha,\alpha;\beta,\beta}  \\
&= \sum_{\alpha,\beta}\left\{
\left(T_{Q}^{X_P} \left\{ \mathbf{1}{\otimes} (U_{Q}^{\dagger})^{n} \right\}\mathbf{1} \right)
(\mathbf{1}\left\{ \mathbf{1}{\otimes} (U_{Q})^{n} \right\}T^{X_P})
\right\}_{\alpha,\alpha;\beta,\beta}  \\
&= \raisebox{-5.0ex}{\includegraphics[scale=0.5]{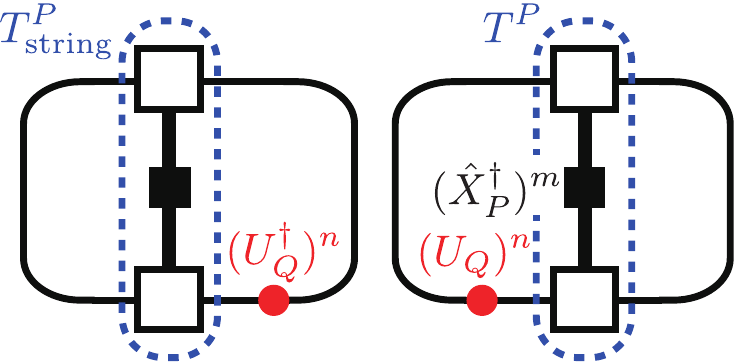}}  
\equiv  \mathcal{O}_{1,\text{L}}(m,n) \mathcal{O}_{1,\text{R}}(m,n) \; ,
\end{split}
\label{eqn:string-in-MPS-SUN}
\end{equation}
where $U_{Q}$ is the projective representation of $Q$ and the transfer matrices are defined as
\begin{equation}
\begin{split}
& [T^{X_P}]_{\bar{\alpha},\alpha;\bar{\beta},\beta} \equiv \sum_{a,b=1}^{d}
\left[A^{\ast}(a)\right]_{\bar{\alpha},\bar{\beta}}
\left[A(b)\right]_{\alpha,\beta} 
\langle a | (\hat{X}^{\dagger}_{P})^{m}| b \rangle \\
& [T_{Q}^{X_P}]_{\bar{\alpha},\alpha;\bar{\beta},\beta} \equiv \sum_{a,b=1}^{d}
\left[A^{\ast}(a)\right]_{\bar{\alpha},\bar{\beta}}
\left[A(b)\right]_{\alpha,\beta} 
\langle a |(\hat{X}_{P})^{m}\hat{Q}^{n} | b\rangle  \; .
\end{split}
\end{equation}
The $\mathbf{V}^{(Q)}_{\text{L},1}$ ($\mathbf{V}^{(Q)}_{\text{R},1}$) denotes the largest left (right) 
eigenvector of the following transfer matrix:
\begin{equation}
[T_{Q}]_{\bar{\alpha},\alpha;\bar{\beta},\beta} \equiv \sum_{a,b=1}^{d}
\left[A^{\ast}(a)\right]_{\bar{\alpha},\bar{\beta}}
\left[A(b)\right]_{\alpha,\beta} 
\langle a |\hat{Q}^{n}|b\rangle   \; .
\end{equation}
Using the properties of the canonical MPS\cite{Garcia-W-S-V-C-08}, we can show that  
the right boundary term $\mathcal{O}_{1,\text{R}}(m,n)$ in Eq.~\eqref{eqn:string-in-MPS-SUN} 
satisfies the following identity\cite{Pollmann-T-12,Hasebe-T-13,Duivenvoorden-Q-ZnxZn-13} 
(see Fig.~\ref{fig:string-boundary-SUN}):
\begin{equation}
\mathcal{O}_{1,\text{R}}(m,n) = 
\omega^{-l(m+n \, n_{\text{top}})} \mathcal{O}_{1,\text{R}}(m,n) \quad 
(l=1,\ldots, N-1) \; .
\end{equation}
That is, if $\omega^{-l(m+n \, n_{\text{top}})} \neq 1$ for some $l$, 
$\mathcal{O}_{1,\text{R}}(m,n)=\mathcal{O}_{1}(m,n)=0$ {\em solely} by symmetry.  
A similar identity is obtained for $\mathcal{O}_{2}(m,n)$ as well.  
Then, these idendity imply that when {\em both} $ \mathcal{O}_{1}(m,n)$ and $ \mathcal{O}_{2}(m,n)$ 
are non-zero, the topological index $n_{\text{top}}$ necessarily satisfies 
\begin{equation}
\omega^{-(m+n \, n_{\text{top}})} = 1 \; .
\label{eqn:string-OP-constraint}
\end{equation}   
For $N=4$, we can use the set of $\mathcal{O}_{1,2}(m,n)$ with 
\begin{equation}
(m,n)=(1,3) \; (\text{class-1)}, \;\;
(2,1) \; (\text{class-2)}, \; \; 
(1,1) \; (\text{class-3)}
\label{eqn:Z4xZ4-string-OP}
\end{equation}
to distinguish between the three topological phases (as well as one trivial one).   
In the SU(4) class-2 phase we discuss here, we expect
\begin{equation}
\begin{split}
& \mathcal{O}_{1,2}(2,1) \neq 0 \, , \\
& \mathcal{O}_{1,2}(1,3) = \mathcal{O}_{1,2}(1,1) = 0 \; .
\end{split}
\end{equation}
In fact, for the solvable SU(4) VBS state discussed in Sec.~\ref{sec:solvable-Ham}, we have 
\begin{equation}
\mathcal{O}_{1,2}(m,n)
= 
\begin{cases}
0                                       & (m,n)=(1,3) \\
1                                       & (m,n)=(2,1) \\
0                                       & (m,n)=(1,1) \; ,
\end{cases}
\end{equation}
which clearly indicate the class-2 topological phase.  

In general, we need a set of $2(N-1)$ SOPs $\mathcal{O}_{1,2}(m,n)$ to identify the PSU($N$) topological 
phases. 
Note that the non-vanishing SOP is the {\em sufficient} condition for the corresponding topological class.   
In other words, 
even if the system is in the topological phase, the corresponding SOP might be zero for some other special reasons. 
%%%%%%%%%%%%%%%%%%%%%%%%%%%%%%%%%%%%%%%%%%%%%%%%%%%%%
%%%%%%%%%%%%%%%%%%%%%%%%%%%%%%%%%%%%%%%%%%%%%%%%%%%%%
\subsection{Reflection}
%%%%%%%%%%%%%%%%%%%%%%%%%%%%%%%%%%%%%%%%%%%%%%%%%%%%%
In contrast to the SU(2) case where the operators $\hat{X}_{P}=S^{z}$ and $\hat{X}_{Q}=S^{x}$  
are hermitian (see Appendix \ref{sec:ZnxZn}), 
$\mathcal{O}_{1,2}(m,n)$ are not invariant under reflection symmetry 
$\mathcal{I}$ (with respect to a site or a bond) for SU($N$) with $N\geq 3$.  
In fact, reflection $\mathcal{I}$ takes them to
\begin{equation}
\mathcal{O}_{1,2}(m,n) \xrightarrow{\mathcal{I}} 
\widetilde{\mathcal{O}}_{1,2}(m,N-n)^{\ast} \; ,
\end{equation}
where $\widetilde{\mathcal{O}}_{1,2}$ here are defined as
\begin{equation}
\begin{split}
& \widetilde{\mathcal{O}}_{1}(m,n) \\
& \equiv 
\lim_{|i-j|\nearrow \infty} 
\Biggl\langle \left\{ \hat{X}_{P}(i) \right\}^{m}  \left\{
\prod_{i < k \leq j } \hat{Q}(k)^{n}
\right\} \left\{\hat{X}^{\dagger}_{P}(j)\right\}^{m} \Biggr\rangle \\
& \widetilde{\mathcal{O}}_{2}(m,n) \\
& \equiv 
\lim_{|i-j|\nearrow \infty} 
\Biggl\langle \left\{ \hat{X}_{Q}(i) \right\}^{m}  \left\{
\prod_{i \leq k  < j} \hat{P}(k)^{n}
\right\} \left\{\hat{X}^{\dagger}_{Q}(j)\right\}^{m} \Biggr\rangle \; .
\end{split}
\end{equation}
The new order parameters $\widetilde{\mathcal{O}}_{1,2}(m,n)$ look similar to 
the original SOP $\mathcal{O}_{1,2}(m,n)$ but are different in the relative position between 
the string and the end points [see Eqs.~\eqref{eqn:def-stringOP-1b} and 
\eqref{eqn:def-stringOP-2b}].   
Now one can repeat the preceding argument [see Eq.~\eqref{eqn:string-in-MPS-SUN} 
and Fig.~\ref{fig:string-boundary-SUN}] on the boundary terms to obtain exactly the same selection 
rule \eqref{eqn:string-OP-constraint}.   
Therefore, one sees that when both $\mathcal{O}_{1}(m,n)$ and $\mathcal{O}_{2}(m,n)$ are 
non-vanishing in a given ground state $|\psi\rangle$, its parity partner $\mathcal{I} |\psi\rangle$ has finite 
$\widetilde{\mathcal{O}}_{1}(m,N-n)$ and $\widetilde{\mathcal{O}}_{2}(m,N-n)$, and hence 
is in another topological phase characterized by $\mathcal{O}_{1,2}(m,N-n)$.   
For instance, the SU(4) class-1 topological phase characterized by $\mathcal{O}_{1,2}(1,3)$ 
is the parity partner of the class-3 phase characterized by $\mathcal{O}_{1,2}(1,4-3)=\mathcal{O}_{1,2}(1,1)$ 
(Fig.~\ref{fig:KT-ZnxZn-string-OP};  
see \href{}{Supplementary Material} for the explicit demonstration).   
%%%%%%%%%%%%%%%%%%%%%%%%%%%%%%%%%%%%%%%%%%%%%%%%%%%%%
\subsection{Numerical results}
%%%%%%%%%%%%%%%%%%%%%%%%%%%%%%%%%%%%%%%%%%%%%%%%%%%%%
To demonstrate the use of the SOP in detecting the SU($N$) topological phases, we plot the value of  
the SOP $\mathcal{O}_{1}(m,n)$ for the model \eqref{eqn:Ha} obtained using iTEBD.   
Note that by the SU(4)-symmetry, we do not need to calculate $\mathcal{O}_{2}(m,n)$.  
That $\mathcal{O}_{1}(2,1)$ is non-vanishing for $\mathcal{H}(a)$ from $a=0$ to $a=1$ gives 
a strong evidence of the class-2 topological phase.   
%%%%%%%%%%%%%%%%%%%%%%%%%%%%%%%%%%%%%%%%%%%%%%%%%%%%%
\begin{figure}[htb]
\centering
\includegraphics[width=0.9\columnwidth,clip]{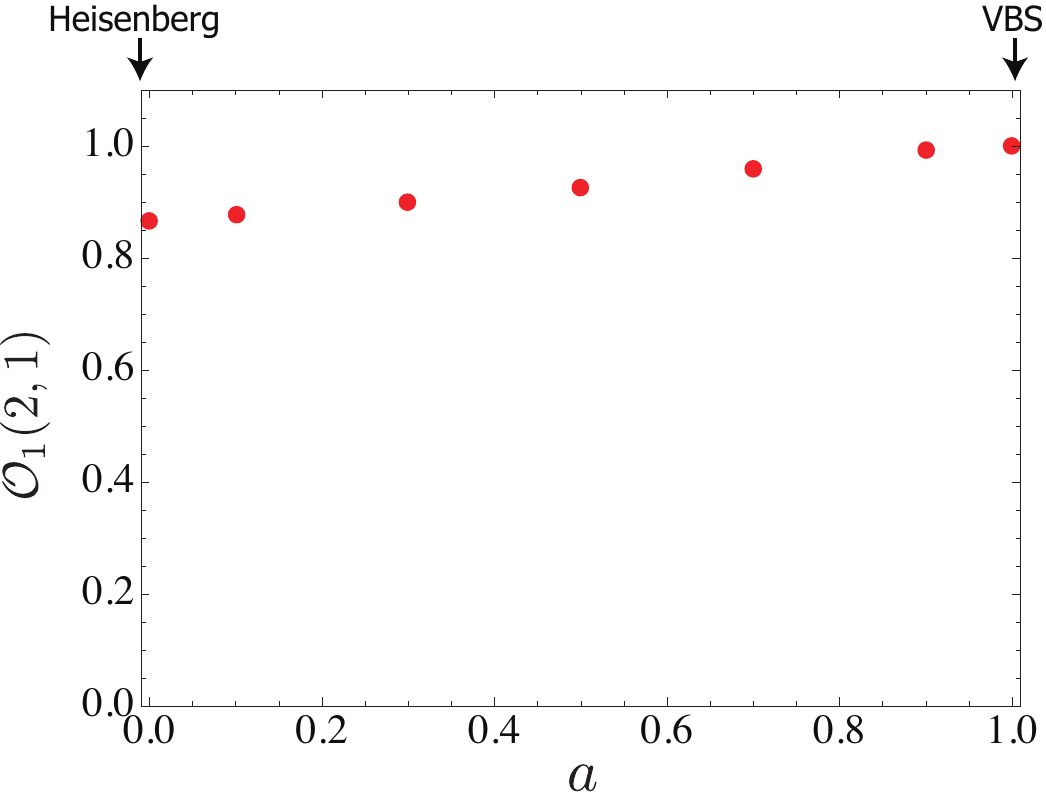}
\caption{(Color online) 
Plot of SOP for $\mathcal{H}(a)$.  $\mathcal{O}_{1}(2,1)$ is non-zero between the Heisenberg point ($a=0$) 
and the VBS point ($a=1$) giving additional evidence for the topological nature. 
\label{fig:string-order-Ha}}
\end{figure}
%%%%%%%%%%%%%%%%%%%%%%%%%%%%%%%%%%%%%%%%%%%%%%%%%%%%%
%%%%%%%%%%%%%%%%%%%%%%%%%%%%%%%%%%%%%%%%%%%%%%%%%%%%%
\subsection{Non-local transformation}
\begin{figure}[htb]
\centering
\includegraphics[width=0.9\columnwidth,clip]{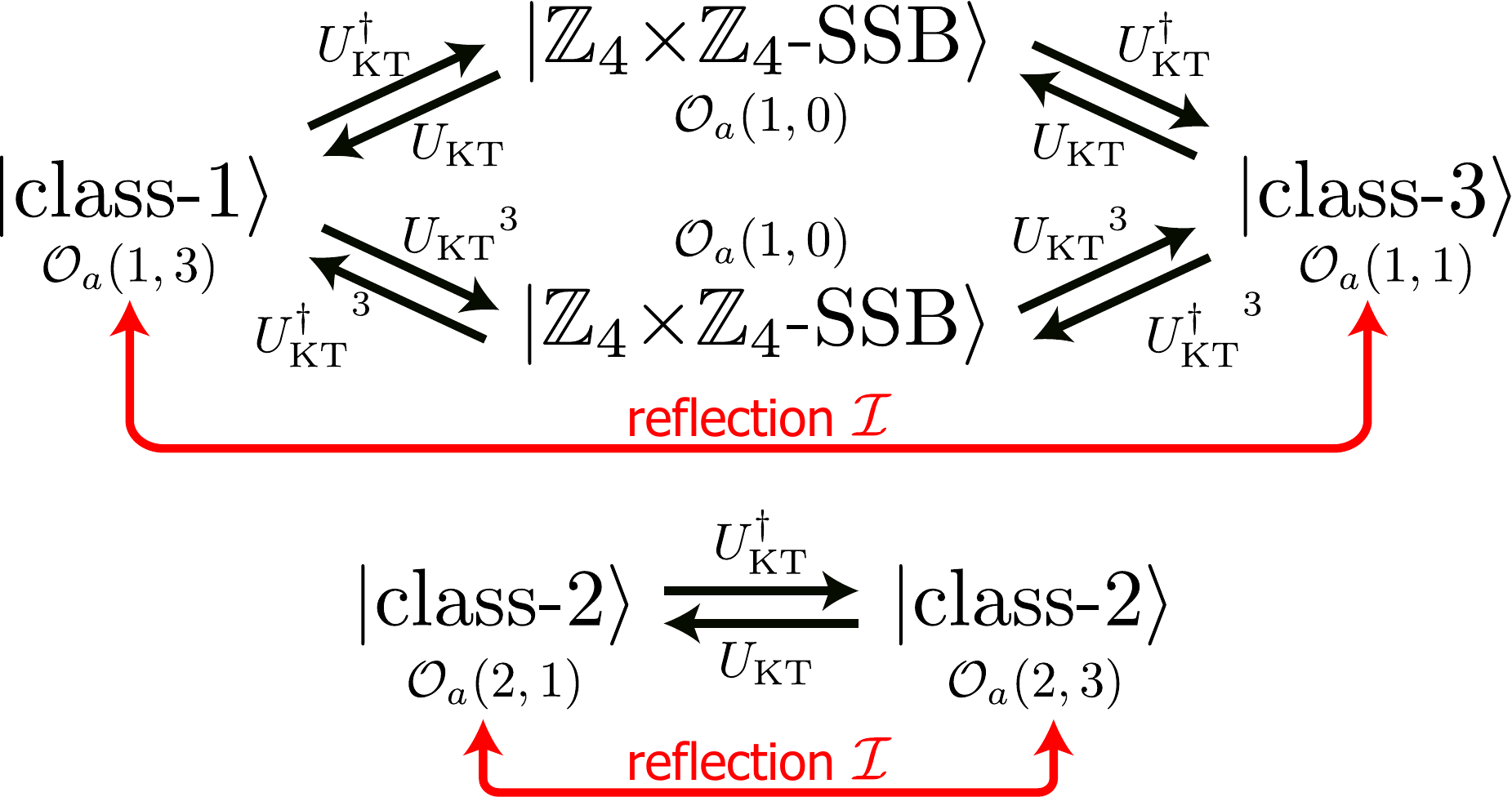}
\caption{(Color online) Generalized Kennedy-Tasaki transformation $U_{\text{KT}}$ and three PSU(4) SPT phases. 
The class-2 state are mapped onto the state of the same topological class by $U_{\text{KT}}$.   
Note that both $\mathcal{O}_{1,2}(2,1)$ and $\mathcal{O}_{1,2}(2,3)$ characterize 
the same class-2 phase [see Eq.~\eqref{eqn:string-OP-constraint}].  
\label{fig:KT-ZnxZn-string-OP}}
\end{figure}

%%%%%%%%%%%%%%%%%%%%%%%%%%%%%%%%%%%%%%%%%%%%%%%%%%%%%
Before concluding this section, we give a remark on the connection between the SOP and the non-local unitary 
transformation (generalized Kennedy-Tasaki transformation) 
eliminating the entanglement of the SPT phase 
that was first introduced in Refs.~\onlinecite{Kennedy-T-92-PRB,Kennedy-T-92-CMP} for the SO(3)-based spin chains 
(see also Refs.~\onlinecite{Okunishi-11,Else-B-D-13} for recent discussions in the context of 
disentangler).  
A straightforward generalization of the above non-local unitary transformation to the PSU($N$) case 
may be given by\cite{Duivenvoorden-Q-ZnxZn-13}
\begin{equation}
U_{\text{KT}} = \exp\left\{
i \frac{2\pi}{N} \sum_{k<j} G_{P}(k) G_{Q}(j) 
\right\} \; .
\label{eqn:def-Kennedy-Tasaki-ZnxZn}
\end{equation}
Then, it is easy to see that the string operators defined in Eqs.~\eqref{eqn:string-op-1} and \eqref{eqn:string-op-2} 
transform (up to phase) as
\begin{equation}
\begin{split}
& U_{\text{KT}}^{\dagger} V_{P}(m,n;i)  U_{\text{KT}} = V_{P}(m,m+n;i) \\
& U_{\text{KT}}^{\dagger} V_{Q}(m,n;i)  U_{\text{KT}} = V_{Q}(m,m+n;i) \; .
\end{split}
\label{eqn:UKT-vs-V}
\end{equation}
This and Eq.~\eqref{eqn:Z4xZ4-string-OP} imply that repeated applications of $U_{\text{KT}}$ 
take the system from one topological phase to another (see Fig.~\ref{fig:KT-ZnxZn-string-OP}).  
In particular, the class-1 and 3 phases can be reduced to conventional phase with 
(spontaneously-broken) local orders, while 
the class-2 is not.  
%%%%%%%%%%%%%%%%%%%%%%%%%%%%%%%%%%%%%%%%%%%%%%%%%%%%%%%
%%%%%%%%%%%%%%%%%%%%%%%%%%%%%%%%%%%%%%%%%%%%%%%%%%%%%%%
%%%%%%%%%%%%%%%%%%%%%%%%%%%%%%%%%%%%%%%%%%%%%%%%%%%%%%%
\section{Symmetry Reduction}
\label{sec:symmetry-reduction}
%%%%%%%%%%%%%%%%%%%%%%%%%%%%%%%%%%%%%%%%%%%%%%%%%%%%%%%
In SPT phases, the list of possible topological phases is closely tied to 
the symmetry we impose on the system, and a phase which is topological under a certain symmetry 
may not be so when we consider a lower symmetry.   
Although the protecting symmetry $\text{PSU($N$)}$ is automatically (i.e., without fine tuning) guaranteed 
almost perfectly in alkaline-earth cold fermions\cite{Gorshkov-et-al-10,Scazza-et-al-14,Zhang-et-al-Sr-14}, 
it would be interesting, from the theoretical point of view, to consider the fate of the topological phases 
when $\text{PSU($N$)}$ gets reduced.  
%%%%%%%%%%%%%%%%%%%%%%%%%%%%%%%%%%%%%%%%%%%%%%%%%%%%%%
\subsection{Systems only with reflection symmetry}
\label{sec:PSUN-to-reflection}
%%%%%%%%%%%%%%%%%%%%%%%%%%%%%%%%%%%%%%%%%%%%%%%%%%%%%%
We begin with the case where the PSU($N$) symmetry is broken down to reflection symmetry  
with respect to the middle of a bond ({\em link-parity} $\mathcal{I}$).   
As is emphasized in Refs.~\onlinecite{Garcia-W-S-V-C-08,Pollmann-T-B-O-10}, 
symmetry operations (whether local or non-local) which keep a given state (which we assume is 
represented as an MPS) invariant are expressed in the form of Eq.~\eqref{eqn:MPS-UAU}:
\begin{equation}
A(m_i) \xrightarrow{\mathcal{I}} A(m_i)^{\text{T}} = 
\be^{i\phi_{\mathcal{I}}} 
U^{\dagger}_{\mathcal{I}} A(m_i) U_{\mathcal{I}} \; ,
\label{eqn:MPS-UAU-inv}
\end{equation}
where $U_{\mathcal{I}}$ satisfies $U_{\mathcal{I}}^{\text{T}} = \pm U_{\mathcal{I}}$.  
Depending on the sign appearing on the right-hand, 
there are two classes for systems with link-parity $\mathcal{I}$ 
(topological when $-1$ and trivial if $+1$)\cite{Pollmann-T-B-O-10}.  

Now let us determine the sign for the SU($N$) VBS state shown in Fig.~\ref{fig:SU(4)VBS}.  
To this end, we first note that the MPS matrices $A(m_i)$ is written as 
\begin{equation}
A(m_i) = \mathcal{R} P(m_i) \; ,
\end{equation}
where $P(m_i)$ is the projection operators from the two fractional objects 
$|\alpha\rangle_{i}$ and $|\beta\rangle_{i} $ [in our SU(4) case they are two {\bf 6} representations 
${\tiny \yng(1,1)}$] 
at site $i$ onto the physical states $|m_i\rangle$:
\begin{equation}
\left[ P(m_i)\right]_{\alpha\beta} \equiv \langle m_i  
|\alpha\rangle_{i}{\otimes}|\beta\rangle_{i}  \; .
\end{equation}
The metric matrix $\mathcal{R}$ creates the SU($N$)-singlet out of
the two fractional objects $|\alpha\rangle_{i}$ and $|\beta\rangle_{i+1}$ on the adjacent site as 
(see Fig.~\ref{fig:SU(4)VBS}):
\begin{equation}
|\text{singlet}\rangle = \mathcal{R}_{\alpha\beta}|\alpha\rangle_{i} |\beta\rangle_{i+1}  \; .
\end{equation}
Then, we can show that $U_{\mathcal{I}}$ is given by the matrix $\mathcal{R}$: 
\begin{equation}
\begin{split}
& \mathcal{R}^{\dagger} A(m_i) \mathcal{R} = 
\mathcal{R}^{\dagger} \left(\mathcal{R}P(m_i)\right) \mathcal{R} \\
&= P(m_i) \mathcal{R} 
= \be^{-i \phi_{I}}\left\{ \mathcal{R} P(m_i) \right\}^{\text{T}} 
= \be^{-i \phi_{I}}A(m_i)^{\text{T}}  \; ,
\end{split}
\end{equation}
where $\phi_{I}$ is 0 when both $P(m_i)$ and 
$\mathcal{R}$ are symmetric/anti-symmetric, and $\pi$ otherwise [in our case, 
$P(m_i)$ are symmetric by construction].  
Therefore, in order to know if $U_{\mathcal{I}}$ is antisymmetric or not, we have only to know how 
the SU($N$)-singlet is constructed out of $|\alpha\rangle_{i}$ and $|\beta\rangle_{i+1}$.   

The SU($N$)-singlet is written as the following fully-antisymmetrized product of $N=2n$ states 
in the fundamental representation $\mathbf{N}$:
\begin{equation}
\begin{split}
& |\text{singlet}\rangle \\
&= \sum_{\{i_k,j_k\}}
\epsilon_{i_1 i_2 \cdots i_n j_1 j_2 \cdots j_n}
|v_{i_1}\rangle  \cdots |v_{i_n}\rangle
|v_{j_1}\rangle \cdots |v_{j_n}\rangle  \\
& = \sum_{\text{partition}}
C_{\{i_k\};\{j_k\}} 
\left\{
\sum_{\{i_k\}}
\epsilon_{i_1 i_2 \cdots i_n}|v_{i_1}\rangle |v_{i_2}\rangle \cdots |v_{i_n}\rangle
\right\} \\
& \times \left\{
\sum_{\{j_k\}}
\epsilon_{j_1 j_2 \cdots j_n}
|v_{j_1}\rangle |v_{j_2}\rangle \cdots |v_{j_n}\rangle 
\right\} \; .
\end{split}
\end{equation} 
As the states inside the braces transform like 
\begin{equation}
\text{\scriptsize $n=N/2$} \left\{ 
\yng(1,1,1)
 \right.  \; ,
\end{equation}
the symmetry of $\mathcal{R}$ is encoded in that of the coefficient $C_{\{i_k\};\{j_k\}}(=\pm 1)$.  
From the antisymmetry of $\epsilon_{i_1 i_2 \cdots i_n j_1 j_2 \cdots j_n}$, one imediatetely sees
\begin{equation}
C_{\{i_k\};\{j_k\}} = (-1)^{n^{2}}C_{\{j_k\};\{i_k\}}  = (-1)^{n}C_{\{j_k\};\{i_k\}} \; .
\end{equation}
Therefore, under the symmtry-lowering perturbation, the class-2 SPT phase 
crosses over to the topological Haldane phase (a trivial phase) when $n=N/2=\text{odd}$ 
(even) (see Fig.~\ref{fig:SUN-topo-phases}).   

%%%%%%%%%%%%%%%%%%%%%%%%%%%%%%%%%%%%%%%%%%%%%%%%%%%%%
\subsection{\texorpdfstring{$\boldsymbol{\mathbb{Z}_{N}{\times}\mathbb{Z}_{N} \mapsto \mathbb{Z}_{2}{\times}\mathbb{Z}_{2}}$}{\mathbb{Z}_{N}{\times}\mathbb{Z}_{N} \mapsto \mathbb{Z}_{2}{\times}\mathbb{Z}_{2}} }
\label{sec:PSUN-to-Z2xZ2}
\begin{figure}[tbh]
\begin{center}
\includegraphics[width=1.0\columnwidth,clip]{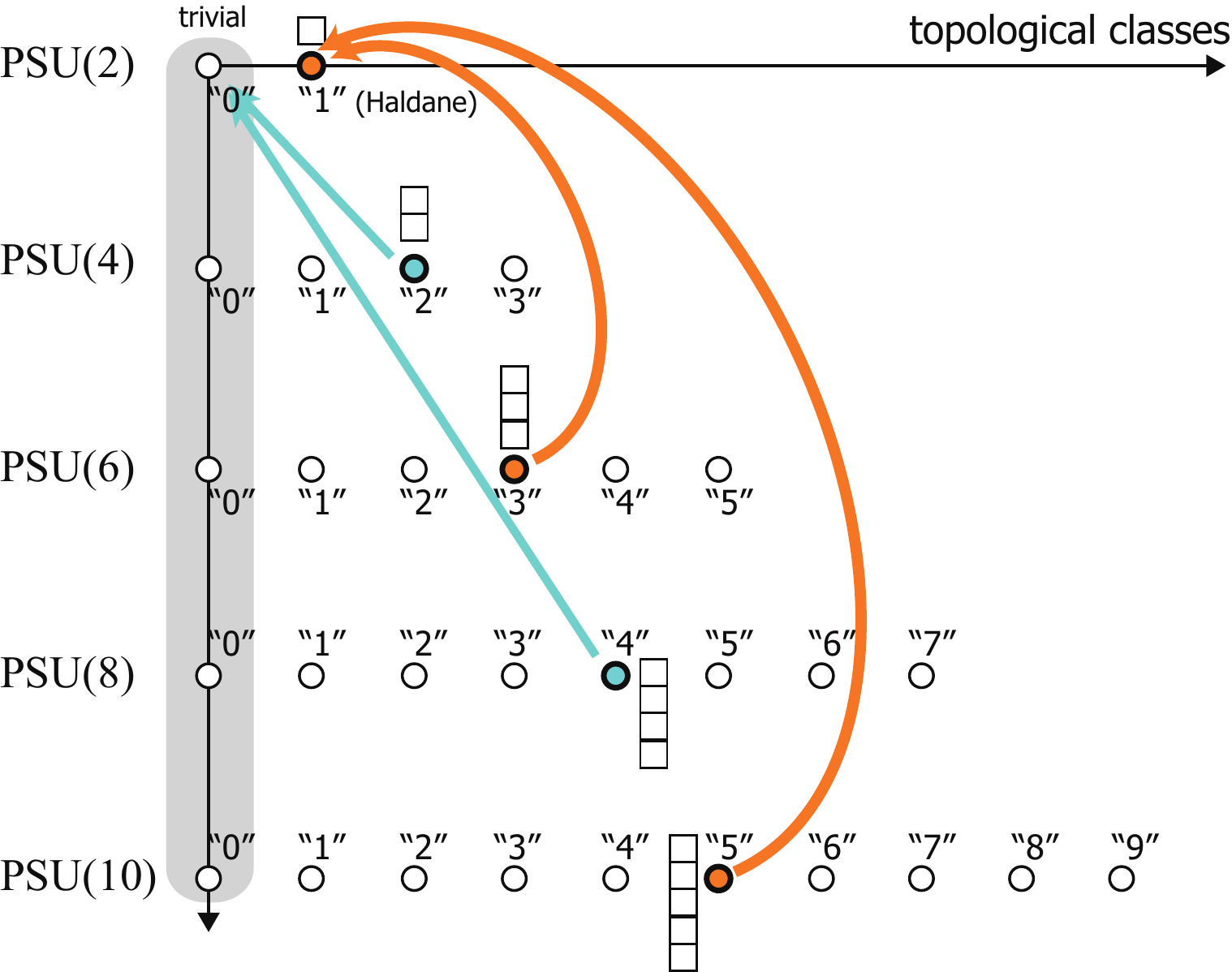}
\end{center}
\caption{(Color online) Fate of SPT phases protected by PSU($2n$) when the symmetry is reduced down to 
link-parity $\mathcal{I}$ (Sec.~\ref{sec:PSUN-to-reflection}) 
or $\mathbb{Z}_{2}{\times}\mathbb{Z}_{2}$ (Sec.~\ref{sec:PSUN-to-Z2xZ2}).  
All these phases are labelled according to the irreducible representation(s) under which 
edge states transform. The label ``$n$'' of the SPT classes stands for the number $n_{\text{Y}}$ 
(see the text) corresponding to the projective representations.  
When only the link-parity or $\mathbb{Z}_{2}{\times}\mathbb{Z}_{2}$ is imposed, only a part of them remains 
topological (warped arrows).  
\label{fig:SUN-topo-phases}}
\end{figure}

%%%%%%%%%%%%%%%%%%%%%%%%%%%%%%%%%%%%%%%%%%%%%%%%%%%%%
As has been seen in Sec.~\ref{sec:def-ZnxZn}, we may regard the $N-1$ PSU($N$) topological phases 
as protected by the subgroup $\mathbb{Z}_{N}{\times}\mathbb{Z}_{N}$ 
(see also Appendix~\ref{sec:PSUN-to-ZnxZn}).    
In that case, the following commutation relation determines the topological classes\cite{Duivenvoorden-Q-ZnxZn-13}:
\begin{equation}
U_P U_Q = \be^{i\frac{2\pi}{N}n_{\text{top}}} U_P U_Q \quad (n_{\text{top}}=0,1,\ldots, N-1) \; .
\label{eqn:U1-U2-exchange-Zn-2}
\end{equation}
As the above $\mathbb{Z}_{N}{\times}\mathbb{Z}_{N}$ contains the 
$\mathbb{Z}_{2}{\times}\mathbb{Z}_{2}$ subgroup generated by $Q^{N/2}$ and $P^{N/2}$ when $N$ is even, 
we may consider the symmetry reduction 
$\mathbb{Z}_{N}{\times}\mathbb{Z}_{N} \mapsto \mathbb{Z}_{2}{\times}\mathbb{Z}_{2}$.  
From the relation
\begin{equation}
\begin{split}
U_Q (U_P)^{N/2} &= \left( \be^{i\frac{2\pi}{N}n_{\text{top}}}\right)^{N/2} (U_P)^{N/2} U_Q   \\
&= (-1)^{n_{\text{top}}} (U_P)^{N/2} U_Q  \; ,
\end{split}
\end{equation}
one can easily see that the projective representations of the two $\mathbb{Z}_{2}$ generators satisfy
\begin{equation}
(U_Q)^{N/2} (U_P)^{N/2} = (-1)^{\frac{1}{2}N n_{\text{top}}} (U_P)^{N/2} (U_Q)^{N/2}  \; .
\end{equation}
It is known \cite{Pollmann-T-B-O-10,Pollmann-B-T-O-12} that in the presence of 
$\mathbb{Z}_{2}{\times}\mathbb{Z}_{2}$-symmetry, the phase is topologically non-trivial when 
the projective representations of the two  $\mathbb{Z}_{2}$s are anti-commuting, 
i.e., $(-1)^{\frac{1}{2}N n_{\text{top}}}=-1$.  
This is possible only when 
\begin{equation}
N=2(2k+1) \;\; (k \in \mathbb{Z}) \quad \text{and} \quad  n_{\text{top}}=\text{odd} \; .
\end{equation}
Since our SU($N$) ($N$: even) topological phase corresponds to $n_{\text{top}}=N/2$, 
it remains topological (i.e., Haldane phase) even after the symmetry gets reduced 
down to $\mathbb{Z}_{2}{\times}\mathbb{Z}_{2}$ 
when $N=2,6,10,\ldots$.   
When $N=0$ (mod 4), on the other hand, the topological phases considered here ($n_{\text{top}}=N/2$) 
smoothly cross over to trivial ones. 
In Fig.~\ref{fig:SUN-topo-phases}, we summarize the crossover predicted here.   
%%%%%%%%%%%%%%%% FIG %%%%%%%%%%%%%%%%%%%%%%%%%%%%%%%%%%%%%
%%%%%%%%%%%%%%%%%%%%%%%%%%%%%%%%%%%%%%%%%%%%%%%%%%%%%%%
%%%%%%%%%%%%%%%%%%%%%%%%%%%%%%%%%%%%%%%%%%%%%%%%%%%%%%%
\section{Conclusion and outlook} 
\label{sec:summary}
%%%%%%%%%%%%%%%%%%%%%%%%%%%%%%%%%%%%%%%%%%%%%%%%%%%%%%%
The possibility of realizing SU($N$) symmetry using alkaline-earth cold atoms provides  
a new arena for the symmetry-protected topological phases. 
In this paper, we have studied the topological properties of the ground state of the SU($N$) Heisenberg 
chain \eqref{eqn:SUN-Heisenberg} with the ``spins'' \eqref{eqn:Young-diagram-GS} at each site, 
especially for $N=4$.  
This model is interesting as it is expected to describe the Mott-insulating region 
of the two-orbital SU($N$) Hubbard model \eqref{eqn:Gorshkov-Ham}.  
From the analysis of the ground state of the solvable VBS Hamiltonian \eqref{eqn:SU4-VBS}, 
we have suspected that the ground state of \eqref{eqn:SUN-Heisenberg} belongs to one of the three topological 
phases predicted for the SU(4)-invariant systems.  To substantiate this, we have calculated the entanglement 
spectrum of an infinite-size system with iTEBD and found that the degeneracy structure is perfectly consistent 
with that expected for the topological class (called {\em class-2} in the text).  
In order to establish the adiabatic continuity between the Heisenberg model and the solvable VBS model, 
we have considered a simple one-parameter deformation $\mathcal{H}(a)$ of the Hamiltonian.  
The entanglement spectrum preserves its degeneracy structure all the way between the two models 
thus establishing the continuity.  

Then, we have investigated how the entanglement spectrum changes when the protecting symmetry 
gets lowered. Specifically, we have considered the situations where the original SU($N$) symmetry 
(which is perfect in the alkaline-earth cold atoms) is reduced to (i) link-parity 
and (ii) $\mathbb{Z}_{2}{\times}\mathbb{Z}_{2}$.  
In both cases, the stability of our topological phase depends on the value of $N$; 
when $N=4n+2$ (i.e., $N=2,6,10,\ldots$), we expect a crossover from our SU($N$) topological phase to 
the Haldane phase.  

Although the entanglement spectrum gives useful insights about the nature of topological phases, 
it may not fully characterize it.  In fact, what is more fundamental is, at least from the group-cohomology 
point of view, the projective representations which is the mathematical representation of the physical 
edge states.  The non-local string-order parameter (SOP) is appealing since it contains the information 
of the projective representation in a manner that may be accessible in experiments.  
We have numerically calculated the SOP $\mathcal{O}_{1}(2,1)$ with iTEBD 
and observed that it stays finite in our topological phase.  
This gives another support to our claim that the ground state of the SU(4) Heisenberg model 
\eqref{eqn:SUN-Heisenberg} is in the class-2 topological phase.  

At least two interesting questions remain to be answered.  
One is about the quantum phase transition(s) out of the topological phase discussed here.  
In fact, an SU($N$) dimerized phase (called ``spin-Peierls'') 
is observed numerically in Ref.~\onlinecite{Bois-C-L-M-T-15} 
next to (i.e., on the smaller-$U$ side of) the SPT phase.   
As the inclusion of higher-order terms in $t/U$ may be mimicked by adding 
terms higher order in $(\mathcal{S}_{i}{\cdot}\mathcal{S}_{i+1})$ to \eqref{eqn:SUN-Heisenberg},  
we may include an extra term that favors dimerization to study the topological-dimerized quantum phase transition.  

Another interesting problem would be the nature of the strong-coupling (Mott) phase of the model 
\eqref{eqn:Gorshkov-Ham} with {\em odd}-$N$.   In this case, the orbital degree of freedom is not 
fully quenched and we obtain an effective Hamiltonian different from \eqref{eqn:SUN-Heisenberg}, 
where the SU($N$) ``spin'' are highly entangled with the orbital degree of freedom\cite{Bolens-C-L-T-15}.  
As the nature of the effective Hamiltonian, which is reminiscent 
of the Kugel-Khomskii-type model\cite{Kugel-K-82}  
for manganese, is not understood, it would be interesting to investigate it by the strategy used here.    
%%%%%%%%%%%%%%%%%%%%%%%%%%%%%%%%%%%%%%%%%%%%%%%%%%%%%%%
\section*{Acknowledgements}
One of the authors (K.T.) has benefitted from stimulating discussions with 
A.~Bolens, S.~Capponi, P.~Lecheminant, and K.~Penc on related projects. 
He was also supported in part by JSPS KAKENHI Grant No.~24540402 and No.~15K05211 
and by the PICS grant from CNRS France.  
%%%%%%%%%%%%%%%%%%%%%%%%%%%%%%%%%%%%%%%%%%%%%%%%%%%%%%%
\appendix
%%%%%%%%%%%%%%%%%%
%%%%%%%%%%%%%%%%%%%%%%%%%%%%%%%%%%%%%%%%%%%%%%%%%%%%%%%
\section{MPS matrices for SU(4) VBS state}
\label{sec:MPS-matrices}
%%%%%%%%%%%%%%%%%%%%%%%%%%%%%%%%%%%%%%%%%%%%%%%%%%%%%%%
In this appendix, we give the matrices necessary for the MPS representation of the SU(4) VBS 
state in Sec.~\ref{sec:solvable-Ham}.   
The MPS for the SU(4) VBS state is given by the following product of six-dimensional matrices 
$A(\mathbf{m}_i) = \Lambda \Gamma(\mathbf{m}_i) = \Gamma(\mathbf{m}_i) \Lambda$ 
(we follow the notations used in Ref.~\onlinecite{Vidal-iTEBD-07}):
\begin{equation}
\begin{split}
& |\text{VBS}\rangle  \\
& =\sum_{\{\mathbf{m}_{i}\}} A(\mathbf{m}_1)A(\mathbf{m}_2) \cdots A(\mathbf{m}_L) 
|\mathbf{m}_1\rangle\otimes |\mathbf{m}_2\rangle \otimes \cdots 
\otimes |\mathbf{m}_L\rangle  \; ,
\end{split}
\end{equation}  
where the summation is taken over all the weights $\mathbf{m}_{i}=(m_{i}^1,m_{i}^2,m_{i}^3)$ 
of the 20-dimensional representation of SU(4) and $\Lambda$ is a diagonal matrix with non-negative diagonal 
elements.   
Throughout this paper, we assume infinite-size systems where the MPS is given by infinite-product 
of matrices $A(\mathbf{m}_{i})$.  For several reasons, it is convenient to use the canonical form of the above MPS, 
where the transfer matrix satisfies certain conditions.   One possible choice of the canonical MPS 
is\footnote{The derivation is sketched in Supplementary Material  
at \href{http://www.example.com/}{http://www.example.com/}.}) 
%%%%%%%%%%%%%%%%%%%%%%%%%%%%%%%%%%%%%%%%%%%%%%%%%%%%%
\begin{widetext}
\begin{equation}
\Lambda = 
\frac{1}{\sqrt{6}} 
\begin{pmatrix}
1 & 0 & 0 & 0 & 0 & 0 \\
0 & 1 & 0 & 0 & 0 & 0 \\
0 & 0 & 1 & 0 & 0 & 0 \\
0 & 0 & 0 & 1 & 0 & 0 \\
0 & 0 & 0 & 0 & 1 & 0 \\
0 & 0 & 0 & 0 & 0 & 1 
\end{pmatrix} 
\end{equation}
%%%%%%%%%%%%%%%%%
\begin{subequations}
\begin{equation}
\Gamma (2,0,0) = \sqrt{6}
\begin{pmatrix}
0 & 0 & 0 & 0 & 0 & 0 \\
 0 & 0 & 0 & 0 & 0 & 0 \\
 0 & 0 & 0 & 0 & 0 & 0 \\
 0 & 0 & 0 & 0 & 0 & 0 \\
 0 & 0 & 0 & 0 & 0 & 0 \\
 1 & 0 & 0 & 0 & 0 & 0 
\end{pmatrix} \; , \;\;
%%%%%%
\Gamma (1,1,0) = \sqrt{3}
\begin{pmatrix}
 0 & 0 & 0 & 0 & 0 & 0 \\
 0 & 0 & 0 & 0 & 0 & 0 \\
 0 & 0 & 0 & 0 & 0 & 0 \\
 0 & 0 & 0 & 0 & 0 & 0 \\
 -1 & 0 & 0 & 0 & 0 & 0 \\
 0 & 1 & 0 & 0 & 0 & 0
\end{pmatrix} \; , \;\;
%%%%%%
\Gamma (1,0,-1) = \sqrt{3}
\begin{pmatrix}
0 & 0 & 0 & 0 & 0 & 0 \\
 0 & 0 & 0 & 0 & 0 & 0 \\
 0 & 0 & 0 & 0 & 0 & 0 \\
1 & 0 & 0 & 0 & 0 & 0 \\
 0 & 0 & 0 & 0 & 0 & 0 \\
 0 & 0 & 1 & 0 & 0 & 0
\end{pmatrix} \; , \;\; 
\end{equation}
\begin{equation}
%%%%%%
\Gamma (0,2,0) = \sqrt{6}
\begin{pmatrix}
0 & 0 & 0 & 0 & 0 & 0 \\
 0 & 0 & 0 & 0 & 0 & 0 \\
 0 & 0 & 0 & 0 & 0 & 0 \\
 0 & 0 & 0 & 0 & 0 & 0 \\
 0 & -1 & 0 & 0 & 0 & 0 \\
 0 & 0 & 0 & 0 & 0 & 0 
\end{pmatrix} \; , \;\;
%%%%%%
\Gamma (1,0,1) = \sqrt{3}
\begin{pmatrix}
 0 & 0 & 0 & 0 & 0 & 0 \\
 0 & 0 & 0 & 0 & 0 & 0 \\
1 & 0 & 0 & 0 & 0 & 0 \\
 0 & 0 & 0 & 0 & 0 & 0 \\
 0 & 0 & 0 & 0 & 0 & 0 \\
 0 & 0 & 0 & 1 & 0 & 0
\end{pmatrix} \; , \;\;
%%%%%%
\Gamma (0,1,-1) = \sqrt{3}
\begin{pmatrix}
0 & 0 & 0 & 0 & 0 & 0 \\
 0 & 0 & 0 & 0 & 0 & 0 \\
 0 & 0 & 0 & 0 & 0 & 0 \\
 0 & 1 & 0 & 0 & 0 & 0 \\
 0 & 0 & -1 & 0 & 0 & 0 \\
 0 & 0 & 0 & 0 & 0 & 0
\end{pmatrix} \; , \;\;  
\end{equation}
\begin{equation}
%%%%%%
\Gamma (1,-1,0) = \sqrt{3}
\begin{pmatrix}
0 & 0 & 0 & 0 & 0 & 0 \\
 -1 & 0 & 0 & 0 & 0 & 0 \\
 0 & 0 & 0 & 0 & 0 & 0 \\
 0 & 0 & 0 & 0 & 0 & 0 \\
 0 & 0 & 0 & 0 & 0 & 0 \\
 0 & 0 & 0 & 0 & 1 & 0 
\end{pmatrix} \; , \;\;
%%%%%%
\Gamma (0,1,1) = \sqrt{3}
\begin{pmatrix}
 0 & 0 & 0 & 0 & 0 & 0 \\
 0 & 0 & 0 & 0 & 0 & 0 \\
 0 & 1 & 0 & 0 & 0 & 0 \\
 0 & 0 & 0 & 0 & 0 & 0 \\
 0 & 0 & 0 & -1 & 0 & 0 \\
 0 & 0 & 0 & 0 & 0 & 0
\end{pmatrix} \; , \;\;
%%%%%%
\Gamma (0,0,-2) = \sqrt{6} 
\begin{pmatrix}
0 & 0 & 0 & 0 & 0 & 0 \\
 0 & 0 & 0 & 0 & 0 & 0 \\
 0 & 0 & 0 & 0 & 0 & 0 \\
 0 & 0 & 1 & 0 & 0 & 0 \\
 0 & 0 & 0 & 0 & 0 & 0 \\
 0 & 0 & 0 & 0 & 0 & 0
\end{pmatrix} \; , \;\;  
\end{equation}
%%%%%%
\begin{equation}
\Gamma (0,0,0)_{\text{A}} = \sqrt{6}
\begin{pmatrix}
0 & 0 & 0 & 0 & 0 & 0 \\
 0 & -\frac{1}{2} & 0 & 0 & 0 & 0 \\
 0 & 0 & \frac{1}{2} & 0 & 0 & 0 \\
 0 & 0 & 0 & \frac{1}{2} & 0 & 0 \\
 0 & 0 & 0 & 0 & -\frac{1}{2} & 0 \\
 0 & 0 & 0 & 0 & 0 & 0
\end{pmatrix} \; , \;\;
%%%%%%
\Gamma (0,0,0)_{\text{B}} =  \sqrt{2} 
\begin{pmatrix}
1 & 0 & 0 & 0 & 0 & 0 \\
 0 & -\frac{1}{2} & 0 & 0 & 0 & 0 \\
 0 & 0 & -\frac{1}{2} & 0 & 0 & 0 \\
 0 & 0 & 0 & -\frac{1}{2} & 0 & 0 \\
 0 & 0 & 0 & 0 & -\frac{1}{2} & 0 \\
 0 & 0 & 0 & 0 & 0 & 1
\end{pmatrix} \; , \;\; 
\end{equation}
\begin{equation}
\Gamma (0,0,2) = \sqrt{6}
\begin{pmatrix}
0 & 0 & 0 & 0 & 0 & 0 \\
 0 & 0 & 0 & 0 & 0 & 0 \\
 0 & 0 & 0 & 1 & 0 & 0 \\
 0 & 0 & 0 & 0 & 0 & 0 \\
 0 & 0 & 0 & 0 & 0 & 0 \\
 0 & 0 & 0 & 0 & 0 & 0
\end{pmatrix} \; , \;\;  
%%%%%%
\Gamma (0,-1,-1) = \sqrt{3}
\begin{pmatrix}
0 & 0 & 0 & 0 & 0 & 0 \\
 0 & 0 & -1 & 0 & 0 & 0 \\
 0 & 0 & 0 & 0 & 0 & 0 \\
 0 & 0 & 0 & 0 & 1 & 0 \\
 0 & 0 & 0 & 0 & 0 & 0 \\
 0 & 0 & 0 & 0 & 0 & 0
\end{pmatrix} \; , \;\;
%%%%%%
\Gamma (-1,1,0) = \sqrt{3}
\begin{pmatrix}
0 & 1 & 0 & 0 & 0 & 0 \\
 0 & 0 & 0 & 0 & 0 & 0 \\
 0 & 0 & 0 & 0 & 0 & 0 \\
 0 & 0 & 0 & 0 & 0 & 0 \\
 0 & 0 & 0 & 0 & 0 & -1 \\
 0 & 0 & 0 & 0 & 0 & 0
\end{pmatrix} \; , \;\; 
\end{equation}
\begin{equation}
%%%%%%
\Gamma (0,-1,1) = \sqrt{3}
\begin{pmatrix}
0 & 0 & 0 & 0 & 0 & 0 \\
 0 & 0 & 0 & -1 & 0 & 0 \\
 0 & 0 & 0 & 0 & 1 & 0 \\
 0 & 0 & 0 & 0 & 0 & 0 \\
 0 & 0 & 0 & 0 & 0 & 0 \\
 0 & 0 & 0 & 0 & 0 & 0
\end{pmatrix} \; , \;\;  
%%%%%%
\Gamma (-1,0,-1) = \sqrt{3}
\begin{pmatrix}
0 & 0 & 1 & 0 & 0 & 0 \\
 0 & 0 & 0 & 0 & 0 & 0 \\
 0 & 0 & 0 & 0 & 0 & 0 \\
 0 & 0 & 0 & 0 & 0 & 1 \\
 0 & 0 & 0 & 0 & 0 & 0 \\
 0 & 0 & 0 & 0 & 0 & 0
\end{pmatrix} \; , \;\;
%%%%%%
\Gamma (0,-2,0) = \sqrt{6}
\begin{pmatrix}
0 & 0 & 0 & 0 & 0 & 0 \\
 0 & 0 & 0 & 0 & -1 & 0 \\
 0 & 0 & 0 & 0 & 0 & 0 \\
 0 & 0 & 0 & 0 & 0 & 0 \\
 0 & 0 & 0 & 0 & 0 & 0 \\
 0 & 0 & 0 & 0 & 0 & 0
\end{pmatrix} \; , \;\; 
\end{equation}
\begin{equation}
%%%%%%
\Gamma (-1,0,1) = \sqrt{3}
\begin{pmatrix}
0 & 0 & 0 & 1 & 0 & 0 \\
 0 & 0 & 0 & 0 & 0 & 0 \\
 0 & 0 & 0 & 0 & 0 & 1 \\
 0 & 0 & 0 & 0 & 0 & 0 \\
 0 & 0 & 0 & 0 & 0 & 0 \\
 0 & 0 & 0 & 0 & 0 & 0
\end{pmatrix} \; , \;\; 
%%%%%%
\Gamma (-1,-1,0) = \sqrt{3} 
\begin{pmatrix}
0 & 0 & 0 & 0 & 1 & 0 \\
 0 & 0 & 0 & 0 & 0 & -1 \\
 0 & 0 & 0 & 0 & 0 & 0 \\
 0 & 0 & 0 & 0 & 0 & 0 \\
 0 & 0 & 0 & 0 & 0 & 0 \\
 0 & 0 & 0 & 0 & 0 & 0
\end{pmatrix} \; , \;\;
%%%%%%
\Gamma (-2,0,0) = \sqrt{6} 
\begin{pmatrix}
0 & 0 & 0 & 0 & 0 & 1 \\
 0 & 0 & 0 & 0 & 0 & 0 \\
 0 & 0 & 0 & 0 & 0 & 0 \\
 0 & 0 & 0 & 0 & 0 & 0 \\
 0 & 0 & 0 & 0 & 0 & 0 \\
 0 & 0 & 0 & 0 & 0 & 0
\end{pmatrix} \; .
\end{equation}
\end{subequations}
Note that the diagonal elements of $\Lambda$ are related to the entanglement spectrum 
$\{\xi_{\alpha}\}$ by $[\Lambda]_{\alpha\alpha}= \be^{-\xi_{\alpha}/2}$.  
\end{widetext}
%%%%%%%%%%%%%%%%%%%%%%%%%%%%%%%%%%%%%%%%%%%%%%%%%%%%%%%
\section{Construction of $\mathbb{Z}_{N}{\times}\mathbb{Z}_{N}$}
\label{sec:ZnxZn}
%%%%%%%%%%%%%%%%%%%%%%%%%%%%%%%%%%%%%%%%%%%%%%%%%%%%%%%
In Sec.~\ref{sec:def-ZnxZn}, we have explicitly constructed the $\mathbb{Z}_{4}{\times}\mathbb{Z}_{4}$ 
subgroup of PSU(4) using the generators of the latter.  
Below, we give the expressions of the $\mathbb{Z}_{N}{\times}\mathbb{Z}_{N}$ generators 
in terms of PSU($N$) for $N=2$ and $3$.   

Regardless of $N$, the first generator $G_{Q}$ is given simply by 
\begin{equation}
G_{Q} = \sum_{k=1}^{N-1} (\vec{\rho})_{k} H_{k} \; ,
\label{eqn:def-G_Q}
\end{equation}
where $H_{k}$ are the $N-1$ Cartan generators and 
$\vec{\rho}$ is the Weyl vector of SU($N$).  The generator $G_{Q}$ 
has the following simple commutation relations with the simple roots $\alpha$:
\begin{equation}
[ G_{Q}, E_{\alpha}]= E_{\alpha} \; , \quad 
[ G_{Q}, E_{-\alpha}]= - E_{-\alpha} \; ,
\end{equation}
which guarantee integer-spaced eigenvalues of $G_{Q}$ 
(for the fundamental representation $\boldsymbol{N}$, they are essentially $1,2,\cdots, N$).  
With this, the first $\mathbb{Z}_{N}$ is generated as
\begin{equation}
Q = c_{N} \exp \left(i\frac{2\pi}{N} G_{Q}\right) \; ,
\end{equation}
where the phase $c_{N}$ has been introduced so that $Q$ satisfy $Q^{N}=1$.  

The expression of the other generator $G_{P}$ depends on $N$.  
For $\mathbb{Z}_{2}{\times}\mathbb{Z}_{2}$, we recover the well-known 
results\cite{Kennedy-T-92-PRB,Kennedy-T-92-CMP}
\begin{subequations}
\begin{align}
& G_{Q} = \rho H =  S^{z} \quad (H=\sqrt{2} S^{z}, \; 
\rho=1/\sqrt{2}) \\
& G_{P} = -\frac{1}{2} E_{\alpha} -\frac{1}{2} E_{-\alpha} = - S^{x}  \; .
\end{align}
The operators $\hat{X}_{P}$ and $\hat{X}_{Q}$ satisfying \eqref{eqn:Z4xZ4-XpXq} are obtained as 
\begin{equation}
\hat{X}_{P} = S^{z} \, , \;\; 
\hat{X}_{Q} = S^{x} \; .
\end{equation}
\end{subequations}

For $\mathbb{Z}_{3}{\times}\mathbb{Z}_{3}$, we have 
\begin{subequations}
\begin{align}
& G_{Q} = \rho_1 H_1 + \rho_2 H_2 =  \sqrt{2} H_1 , \; 
(\vec{\rho}=(\sqrt{2},0) ) \\
& G_{P} = 
-\frac{i}{\sqrt{3}}\sum_{k=1}^{3}(E_{\alpha_k} - E_{- \alpha_k}) \; ,
\end{align}
\end{subequations}
where $\alpha_{1,2}$ are the simple roots of SU(3) and $\alpha_{3}$ is defined by  
$\alpha_{3}\equiv -\alpha_1 - \alpha_2$.   
The operators $X_{P}$ and $X_{Q}$ satisfying Eq.~\eqref{eqn:Z4xZ4-XpXq} are given by
\begin{subequations}
\begin{align}
& X_{P} = H_1 - i H_2 \\
& X_{Q} 
= \sqrt{\frac{2}{3}}\left\{ 
E_{-\alpha_1} + E_{-\alpha_2} + E_{-\alpha_3} 
\right\}   \quad (\alpha_3 \equiv -\alpha_1 - \alpha_2)  \; .
\end{align}
 \end{subequations}
 %%%%%%%%%%%%%%%%%%%%%%%%%%%%%%%%%%%%%%%%%%%%%%%%%%%%%%
\section{$\text{PSU(\texorpdfstring{$\boldsymbol{N}$}{N})}$ and $\mathbb{Z}_{N}{\times}\mathbb{Z}_{N}$}
\label{sec:PSUN-to-ZnxZn}
%%%%%%%%%%%%%%%%%%%%%%%%%%%%%%%%%%%%%%%%%%%%%%%%%%%%%%
Since PSU($N$) and $\mathbb{Z}_{N}{\times}\mathbb{Z}_{N} [\subset \text{PSU($N$)}]$ 
share the same cohomology 
group $H^{2}(\text{PSU($N$)},\text{U(1)})=H^{2}(\mathbb{Z}_{N}{\times}\mathbb{Z}_{N},\text{U(1)})=\mathbb{Z}_{N}$, 
a phase which is topological under PSU($N$) may remain so even if we weakly break \text{PSU($N$)} 
down to $\mathbb{Z}_{N}{\times}\mathbb{Z}_{N}$.   
As we have seen in Sec.~\ref{sec:SPT}, 
when the system has the full PSU($N$)-symmetry, the entanglement spectrum exhibits the degeneracy pattern 
that is compatible with SU($N$)-symmetry.  
That is, the degeneracy of each entanglement level should find the corresponding entry in TABLE \ref{tab:Young}.  
Now let us consider how the reduction of the symmetry down to a subgroup 
$\mathbb{Z}_{N}{\times}\mathbb{Z}_{N}$ changes the entanglement spectrum.  

As the unitary matrices $U_{P,Q}$ assume block-diagonal forms reflecting  
the structure of the entanglement levels, 
the relation \eqref{eqn:U1-U2-exchange-Zn} holds for each block 
corresponding to the degenerate entanglement levels $\lambda$ 
\begin{equation}
U_{P}(\lambda)U_{P}(\lambda) =  \be^{i\Phi_{QP}} U_{Q}(\lambda)U_{P}(\lambda) 
= \be^{i \frac{2\pi}{N}n_{\text{top}}} U_{Q}(\lambda)U_{P}(\lambda) \; .
\end{equation}
This restricts the degree of degeneracy $D_{\lambda}$ of each entanglement level\cite{Bois-C-L-M-T-15}.   
Calculating the determinant of both sides of the above equation, one obtains
\begin{equation}
\begin{split}
\text{det}\, (U_{P}(\lambda)U_{Q}(\lambda)) &= \text{det} \,U_{P}(\lambda)\text{det}\,U_{Q}(\lambda)  \\
& = (\be^{i \frac{2\pi}{N}n_{\text{top}}})^{D_{\lambda}} \text{det}\,U_{P}(\lambda)\text{det}\,U_{Q}(\lambda) \; ,
\end{split}
\label{eqn:det-U1-U2-Zn}
\end{equation}
which immediately implies $(\be^{i \frac{2\pi}{N}n_{\text{top}}})^{D_{\lambda}}=1$.   
When $N$ and $n_{\text{top}}$ are mutually co-prime, $D_{\lambda}$ should be 
integer multiple of $N$.  Otherwise, $D_{\lambda}$ of each level may be smaller.  
In particular, the entanglement spectrum of the class-1 PST phase exhibits 
the $N$-fold degenerate structure for any $N(\geq 2)$, which is consistent with 
the results of the explicit calculation\cite{Katsura-H-K-08} 
for the SU($N$) VBS chain based on another representation\cite{Affleck-K-L-T-87,Affleck-K-L-T-88}.  

For $N=4$ ($\mathbb{Z}_{4}{\times}\mathbb{Z}_{4}$), there are three topological phases 
(i) class-1 ($n_{\text{top}}=1$), (ii) class-2 ($n_{\text{top}}=2$), and (iii) class-3 ($n_{\text{top}}=3$).   
In the class-1 and 3 phases, $D_{\lambda}=0$ (mod 4), while {\em any even integers} are 
allowed for $D_{\lambda}$ in the class-2 phase.  
Therefore, the degeneracy pattern observed in Sec.~\ref{sec:ES} in general may be 
modified when we relax the full PSU(4) symmetry down to $\mathbb{Z}_{4}{\times}\mathbb{Z}_{4}$, 
although the system still stays in the same phase.  
For instance, the lowest six-fold-degenerate level might be split into, e.g., three two-fold-degenerate levels.  

For instance, we may add to the original Hamiltonian \eqref{eqn:SUN-Heisenberg} 
the following $\mathbb{Z}_{N}{\times}\mathbb{Z}_{N}$-invariant perturbation 
[see Eq.~\eqref{eqn:Z4xZ4-XpXq}]
\begin{equation}
\mathcal{V}_{N}
= g_{N} \sum_{i} \left\{ 
\left( \hat{X}_{P}(i) \right)^{N} + \left( {\hat{X}_{P}}^{\dagger}(i) \right)^{N} 
\right\} \; , 
\end{equation}
which is a generalization of the well-known single-ion anisotropy $D\sum_{i}(S_{i}^{z})^{2}$ 
in the usual spin chains [note $\hat{X}_{P}(i)={\hat{X}_{P}}^{\dagger}(i)=S^{z}$ for $N=2$].    

%%%%%%%%% BIB-files %%%%%%%%%%%%%%%%%%%%%%%%%%%%%%%%%%%%%%%%
\bibliographystyle{apsrev4-1}
%\bibliography{../references/alkaline-earth}
%%%%%%%%%%%%%%%%%%%%%%%%%%%%%%%%%%%%%%%%%%%%%%%%%%%%%%
\end{document}